\documentstyle[aaspp4]{article}

\newcommand{\D}{\partial}

\def\dsigw{\mbox{$\dot\Sigma_{\rm w}$}{}}
\def\msun{{\,M_\odot}}
\def\simlt{\lower.5ex\hbox{$\; \buildrel < \over \sim \;$}}
\def\simgt{\lower.5ex\hbox{$\; \buildrel > \over \sim \;$}}
\def\kms{{\rm\,km\,s^{-1}}}
\def\gscm2{{\rm\,g\,s^{-1}\,cm^{-2}}}
\def\cm{{\rm\,cm}}
\def\ncm3{{\rm\,cm^{-3}}}
\def\gms{{\rm\,g\,s^{-1}}}
\def\pc{{\rm\,pc}}
\def\gcm3{{\rm\,g\,cm^{-3}}}
\def\refbook#1{\refindent#1}
\def\refindent{\par\noindent\hangindent=3pc\hangafter=1 }
\def\aa#1#2#3{\refindent#1, A\&A, {\bf#2}, #3.}
\def\aalett#1#2#3{\refindent#1, A\&A {\it (Letters)}, {\bf#2}, #3.}

\def\apj#1#2#3{\refindent#1, {\it ApJ}, {\bf#2}, #3.}
\def\apjlett#1#2#3{\refindent#1, {\it ApJ (Letters)}, {\bf #2}, #3.}
\def\apjsup#1#2#3{\refindent#1, ApJS, #2, #3}

\def\baas#1#2#3{\refindent#1, BAAS, #2, #3}

\def\mnras#1#2#3{\refindent#1, {\it MNRAS}, {\bf#2}, #3.}
\def\nature#1#2#3{\refindent#1, {\it Nature}, {\bf #2}, #3.}

\begin{document}
\title{Accretion Disk Evolution with Wind Infall II.  \\Results
of 3D Hydrodynamical Simulations with an Illustrative Application to Sgr A*}
\author{Robert Coker\altaffilmark{1}}
\affil{Department of Physics, University of
Arizona, Tucson, AZ 85721 (rfc@physics.arizona.edu)}
\author{Fulvio Melia\altaffilmark{2}}
\affil{Department of Physics \& Steward Observatory, University of
Arizona, Tucson, AZ 85721 (melia@as.arizona.edu)}
\author{Heino Falcke}
\affil{Max-Planck-Institut f{$\ddot{\rm u}$}r Radioastronomie,
Auf dem H{$\ddot{\rm u}$}gel 69, D-53121, Bonn, Germany. (hfalcke@mpifr-bonn.mpg.de)}
\affil{Steward Observatory, University of Arizona, Tucson, AZ 85721}
\altaffiltext{1}{NASA GSRP Fellow.}
\altaffiltext{2}{Presidential Young Investigator.}

\begin{abstract}
In the first paper of this series, using analytic tools, we examined
how the evolution and structure of a massive accretion disk may be
influenced significantly by the deposition of mass and angular
momentum by an infalling Bondi-Hoyle wind.  Such a mass influx impacts
the long-term behavior of the disk by providing additional sources of
viscosity and heating.  In this paper, we make a significant
improvement over this earlier work by incorporating the results of 3D
hydrodynamical simulations of the large scale accretion from an
ambient medium into the disk evolution equations developed previously.
We discuss in detail two models, one with the axis of the disk
parallel to, and the second with the axis oriented perpendicular to
the large scale Bondi-Hoyle flow.  We find that the mass inflow rate onto
the disk within logarithmic annuli is roughly constant with radius and
that the impacting wind carries much less specific 
angular momentum than Keplerian.  We also find, in general, that
the infrared spectrum of a wind-fed disk system is steeper than that
of a Shakura-Sunyaev configuration, due mainly to the dissipation
of the wind's kinetic energy at the disk's surface.
In applying our results to the Galactic Center black hole candidate,
Sgr A*, accreting from nearby stellar winds, we demonstrate that a
high wind inflow rate of order $10^{-4}M_\odot$ per year cannot be
incorporated into a fossil disk without a significant dissipation of
kinetic energy at all radii. Such a high dissipation would violate
current infrared and near-infrared limits on the observed
spectrum of Sgr A*.
\end{abstract}


\section{Introduction}
In this paper, we will continue our study of the effects on massive
accretion disk evolution due to the continued infall from
Bondi-Hoyle accretion. This problem is of general interest to
systems in which the mass influx is not planar at the disk's outer
edge, but rather originates from a 3-dimensional distribution of wind
sources.  An example is the formation and evolution of a protostellar
disk, or a disk embedded within a cluster of stellar wind sources,
such as may be the case for the black hole candidate Sgr A* at the
Galactic Center.  In our application, 3D hydrodynamic simulations of
the Bondi-Hoyle infall are used to provide more realistic profiles of
the accreted mass, temperature, and specific angular momentum as
functions of radius.  These profiles are incorporated into the
one-dimensional equations for the temporal evolution of
accretion disks with wind infall developed in \cite{FM97} (Paper I).

\subsection{Galactic Center Bondi-Hoyle Accretion}
Spectral and kinematic studies suggest that the powerful nonthermal
radio source Sgr A*, located at the center of the Milky Way, is a
supermassive compact object with a mass $\sim 2-3\times{10}^6\msun$.
This inference is based on the large proper motion of nearby stars
(\cite{HA95}; \cite{EG97}; \cite{GEOE97}),
the spectrum of Sgr A* (e.g., \cite{MJN92}), 
its low proper motion ($\simlt 20 \kms$; \cite{B96}),
and its unique location (\cite{LAS91}).
Measurements of high outflow velocities
associated with sources near Sgr A* (\cite{K91}; \cite{G91}),
as well as emission line (\cite{G86}; \cite{HKS82}; \cite{AHH90}; \cite{G91})
and radio continuum observations (\cite{YM91}),
provide clear evidence of a hypersonic wind pervading the inner
parsec of the Galaxy.  This Mach $=10-30$ wind has a velocity
$v_w=500-1000\; \kms$, a number density $n_w=10^{3-4}\;\ncm3$, and a total
mass outflow rate $\dot M_w=3-4\times10^{-3}\;M_\odot$ yr$^{-1}$.

Sgr A*'s radiative characteristics, should be directly or indirectly
due to its accretion of this wind.  In the classical Bondi-Hoyle (BH)
scenario (\cite{BH44}), the mass accretion rate for a uniform
hypersonic flow is
\begin{equation}\label{mdot}
\dot M_{BH} = \pi {R_A}^2 m_H n_w v_w,
\end{equation}
in terms of the accretion radius $R_A \equiv 2 G M / {v_w}^2$.
At the Galactic Center, using $M \sim 10^6 \msun$,
$n_w \sim 5.5 \times 10^3 \ncm3$,
and $v_w \sim700 \kms$, we would therefore expect an accretion rate
$\dot M_{BH} \sim 6 \times 10^{21} \gms$ onto the black hole, with a capture
radius $R_A \sim 0.02 \pc$.  Since this accretion rate is sub-Eddington
for a million solar mass object, the accreting gas is mostly
unimpeded by the escaping radiation field and is thus
essentially in hydrodynamic free-fall starting at $R_A$.
Our initial numerical simulations of this process, assuming a highly
simplistic uniform flow past a 
$1\times{10}^6\msun$ point mass (\cite{RM94}; \cite{CM96}) have
verified these expectations.

Whether or not this accretion
is mediated via an accretion disk or it 
constitutes a more or less quasi-spherical
infall depends critically on the specific angular momentum carried by
the captured gas.  Therefore, it is not always clear whether disk or
quasi-spherical accretion dominates or if both are present.
However, in the case of Sgr A*, if the surrounding winds are relatively
uniform, the fluctuations in the accreted specific angular momentum are
sufficiently small that only a small accretion disk, less than about $\sim 50$
$R_s$ in radius (where $R_s \equiv 2 G M / c^2$ is the Schwarzschild radius)
is then expected to form (\cite{RM94}; \cite{CM96}).  However, non-uniformities 
in the ambient medium, due to individual stellar
wind sources, could result in a greater specific angular
momentum being accreted with the captured material (\cite{CM97}), and 
hence account for a larger circularization radius (but still $\la 5000 R_s$)
and presumably a larger and brighter accretion disk. 
Another way of saying this is that the 
infalling gas from a non-uniform medium may retain 
a larger Keplerian energy that must be dissipated 
(or, possibly, advected) as it drifts 
inwards towards the black hole.

\subsection{Massive Accretion Disk Models}
Although the broadband radiative emission from Sgr A* may be
produced either in the quasi-spherical accretion portion of the
inflow (\cite{M94}) or by a radio jet
(\cite{FMB93}, \cite{FB99}), the low
total luminosity of Sgr A* ($\simlt$ a few $10^5 L_{\sun}$;
\cite{Z95}) seems to point to either a
much lower accretion rate ($\ll 10^{21} \gms$), if
one assumes a conversion efficiency
of $10\%$ rest mass energy into radiation, or
a low conversion efficiency ($\simlt 10^{-5}$).
Sgr A* appears to be the ideal case for applying the concept
of {\sl any} large, stable accretion disk fed by quasi-spherical accretion since the
disk acts as a resevoir for the infalling
plasma, thereby decreasing the expected conversion efficiency and luminosity.

An $\alpha$-disk model with an accretion rate implied by Equation (1)
is not consistent with the observed infrared faintness of Sgr A*
(Melia 1994).  Thus,
if a disk is present in this source, it must either be very faint due
to its ``fossilized'' nature, or it must be advection dominated, such that
most of its dissipated energy is swept through the event horizon before
it can be radiated away.  The disk evolution equations derived in
Paper I (\cite{FM97}) apply to the former case.  The structure of 
advection dominated disks, on the other hand, depends on both the
state variables and their gradients, so their evolution must be
handled globally, not locally.  Thus, although we discuss both in this
paper, our quantitative results for the impact of a wind infall on
the disk's evolution, apply primarily to the former.

\subsubsection{A Fossil Disk}
It has been suggested (\cite{FH94}, \cite{F96}) that Sgr A* might be surrounded by a
large, slowly accreting, fossil disk, perhaps formed from the
remnants of a tidally-disrupted star that ventured too close to
the black hole.  Depending on the strength of such a disruption,
as much as half of the stellar mass may be left as a remnant
surrounding the black hole (\cite{KM96} and references cited
therein).  This remnant may, under some circumstances, form a relatively
cold disk that evolves without a substantial mass infusion at its
outer edge.  A fossil disk may also exist as the remnant of a phase of higher
activity in the past (\cite{FH94}). In both cases, one expects 
that the large-scale Bondi-Hoyle inflow must be captured and incorporated into
the fossil disk.  The resulting
observational signature will then depend strongly on the ratio of wind
to disk accretion rates, the viscosity in the disk, the specific angular
momentum of the infalling matter and the time scales involved.
For example, the fossil disk could remain faint for long periods of
time if the inflowing gas has a large specific angular momentum, for it
would then get absorbed onto the plane at large radii where the dissipation
rate is relatively small (Paper I).

\subsubsection{An Advection Dominated Disk}
The disk in sources such as Sgr A* may also be faint because it is 
advection dominated 
(\cite{A95}; \cite{NYM95}).  In this case, the accretion is highly advective, 
decreasing the expected luminosity by up to an order of magnitude or more.
This occurs because a large fraction ($90\%$ or more) of the dissipated
gravitational energy flux is carried inwards through the event 
horizon by the disk plasma.
This model requires a small mass accretion rate ($\simlt 
10^{-5}\;M_\odot$ yr$^{-1}$)
compared to the value indicated by Equation (1), a 
large $\alpha$ ($\simgt 0.1$), 
and a large disk scale size ($\sim 0.02 \pc$) in order to produce the 
required low efficiency.  Some of the results discussed in this paper,
particularly the kinetic energy dissipated and radiated as the infalling
wind impacts the plane, will be valid in a qualitative sense
for this type of disk just as they are for a fossilized structure.
We caution, however, that the evolution of an advection dominated
disk cannot be described adequately with the equations presented below,
and so quantitative conclusions regarding the impact of a wind
infall must in this case await future work. 

\subsection{Scope and Outline of the Paper}
In Section 2 we summarize the pertinent results of Paper I and
discuss how the numerical results will be incorporated into the model.
In Section 3 we present some of the details and results of the 
hydrodynamical simulations.  The implications for the Galactic Center
are discussed in Section 4.  We provide a summary 
and describe further applications of our model in Section 5.

\section{The Disk Equations with Infall}

The structure of the infalling wind is determined by the 
hydrodynamical simulations (see Section 3).  This wind impacts
the disk, resulting in a total dissipation rate (for each side
of the disk):
\begin{equation}
D(r,t)=D_{\rm d}+D_{\rm w},
\end{equation}
where $D_{\rm d}$ is the internal energy dissipation caused
by turbulent viscosity and differential rotation of the disk
and is given by:
\begin{equation}\label{Dd}
D_{\rm d}={1\over2}\,\nu\Sigma\left(r{\D\omega\over\D r}\right)^2,
\end{equation}
and $D_{\rm w}$ is the dissipation rate due to the infalling wind's
kinetic energy relative to the disk being converted into thermal energy:
\begin{equation}
D_{\rm w}= {1\over2}\dsigw {v_w}^2,
\end{equation}
where $v_w$ is the wind velocity relative to the disk's motion and,
assuming $v^d_r \ll v_r$ (see Eq.~\ref{vr}), is given by:
\begin{equation}
v_w = r^2 \omega^2 (1-\xi)^2 + {v_z}^2 + {v_r}^2.
\end{equation}
In the preceding equations, $\Sigma$ is the disk's surface density, $r$ is the radius,
$\dsigw$ is the mass infall onto the disk per unit area and time, $\omega$ is the disk's rotational
frequency, ${\D\omega/\D r}$ is the radial derivative of $\omega$, $v_z$ and $v_r$
are the z- and r-component, respectively, of the 
wind's velocity just before it impacts the disk, $\nu$ is the internal disk viscosity, 
and $\xi$ is the ratio of the impacting wind's azimuthal velocity to its Keplerian value.
It is then assumed that the total observed energy flux is the sum of the energy dissipated
in the disk and the energy dissipated by the impact of the wind on
the surface of the disk so that we have
\begin{equation}
F_{\rm tot}=\sigma T_{\rm eff}(r,t)^4 = D(r,t),
\end{equation}
where $\sigma$ is the Stefan-Boltzmann constant.
This expression assumes that the disk is optically thick.
To obtain the spectrum, we use a sum of blackbodies
at the local effective temperature $T_{\rm eff}(r,t)$.

>From the hydrodynamical simulations, we get spatial and temporal profiles for
$\xi$, $\dsigw$, and $v_w$, while the assumption of a Keplerian disk provides 
$\omega$ and ${\D\omega/\D r}$.  This leaves $\Sigma$ and $\nu$ to be
determined.  For $\Sigma$, we use the well-known differential equation for the
time evolution of the surface density of an accretion disk, but now including
the effects of wind infall:
\begin{equation}\label{dsigw}
{\D \Sigma\over\D t}={3\over r} {\D\over\D r}\left[
r^{1/2}{\D\over\D r}\left(\nu\Sigma r^{1/2}\right) + {2\over3}(1-\xi) r^2 \dsigw\right]
+\dsigw\,.
\end{equation}
For $\nu$, we adopt the standard Shakura-Sunyaev $\alpha$-prescription so that:
\begin{equation}
\nu=\alpha c_{\rm s} z_0\;,
\end{equation}
where $\alpha$ is the dimensionaless viscosity parameter,
$c_{\rm s}$ is the local sound speed in the midplane of the disk given by 
\begin{equation}
c_{\rm s}=\sqrt{{\gamma k_{\rm b}T(0)}\over{\mu m_{\rm p}}}, 
\end{equation}
and $z_0$ is the scale height of a gas-pressure dominated configuration, given by 
\begin{equation}
z_0 = {c_{\rm s}\over\omega}\left\{\left[\left({1\over2}{{\dsigw}\over{\Sigma\omega}}{{v_{\rm z}}\over{c_{\rm s}}}\right)^2
+1\right]^{1/2}-~{1\over2}{{\dsigw}\over{\Sigma\omega}}{{v_{\rm z}}\over{c_{\rm s}}}\right\}.
\end{equation}
As discussed in Paper I, the temperature, $T(0)$, at the midplane of the disk is given by
\begin{equation}\label{T0}
T(0)^4={1\over\sigma}\left[\left({{3\kappa\Sigma}\over4}+1\right)D_{\rm d}+D_{\rm w}\right].
\end{equation}
We assume the opacity, $\kappa$, is due to free-free scattering and is given by
\begin{equation}
\kappa = 6.6\times10^{22}\;{\rm cm}^2\;{\rm g}^{-1} \left({\rho\over{{\rm 1}~\gcm3}}\right) 
\left({T\over{\rm 1\;K}}\right)^{-7/2}.
\end{equation}

Thus, for a given initial set up, one starts with an estimate of $T(0)$ in order
to obtain $c_{\rm s}$, $\nu$, and $D(r)$.  Using Equation~\ref{T0}, a new 
temperature estimate is made; the system rapidly converges.  
Since $c_{\rm s}$ changes slowly, the sound speed from the previous
time step can be used in a Runge-Kutta algorithm with adaptive step size
for the time evolution of $\Sigma(r,t)$.

It is also useful to know the rate, $\dot M$, with which the mass is flowing radially 
through the disk:
\begin{equation}
\dot M = 2\pi r v^d_r \Sigma\;,
\end{equation}
where $v^d_r$ is the radial velocity of gas in the disk and is
given by the expression
\begin{equation}\label{vr}
v^d_r=-{3\over r\Sigma}\left[r^{1/2}{\D\over\D r}\left(\nu \Sigma
r^{1/2}\right)+ {2\over3} (1-\xi) r^2 \dsigw\right].
\end{equation}

\section{The Hydrodynamical Simulations}
\subsection{The Code}
We use a modified version of the 3D hydrodynamics 
code ZEUS, a general purpose code for MHD fluids developed at 
NCSA (\cite{SN92}; \cite{No94}).
The code uses Eulerian finite differencing with the following 
relevant characteristics:
fully explicit in time; operator and 
directional splitting of the hydrodynamical
variables; fully staggered grid; second-order (van Leer) upwinded, monotonic
interpolation for advection; consistent advection to evolve internal
energy and momenta; and explicit solution of internal energy.  
More details can be found in the references.

We concentrate primarily on the hydrodynamical aspects of the infall and therefore exclude 
the effects of magnetic heating and radiative cooling, even though bremsstrahlung and, especially,
magnetic bremsstrahlung emission may become significant at smaller radii.
Thus, the gas is assumed to be an adiabatic, polytropic gas with {$\gamma=5/3$}.
The quantitative results presented here may be altered with the inclusion of more of the relevant physics (heating and cooling,
magnetic fields, and individual wind sources); this may be 
addressed in future work.

\subsection{The Setup}
The first step is to run a large scale simulation, with a volume of solution
$16 R_A$ or $\sim$0.28$\pc$ on a side; the results are used as
boundary conditions for subsequent small scale simulations which are 0.001$\pc$ on a side.
The size of the large simulation was chosen so as to maximize spatial resolution
while minimizing boundary effects.  The size of the small simulation was chosen
to again maximize resolution while having its outer boundary as close as possible to the 
large simulation's inner boundary.  The large scale run lasts 1500 years to
ensure that equilibrium is reached and the last 100 years are used for the 100 year long small
scale simulations.  A single simulation, combinding the large and small scale,
would be ideal, but the high spatial resolution required in the central region, 
when combined with the need to satisfy the Courant condition, makes such a calculation 
computationally prohibitive at present.

Since we are primarily interested in a basic understanding of the disk evolution
due to the infalling wind, we use a simple hydrodynamical setup.  That is,
we assume the wind is uniform and enters the volume of solution 
from the +z directon; all other faces of the volume of solution have outflow boundary
conditions.  The total flow into the large volume is $3\times10^{-3}\;M_\odot$ yr$^{-1}$ with
a velocity of $700 \kms$, a Mach number of 10, and number density of $5.5\times10^3 \ncm3$,
consistent with the conditions in the Galactic Center.  
For all simulations, a point mass of $1\times{10}^6\msun$ is used; these
simulations were initiated before the more reliable value of $2.6\times 10^6\msun$
was known precisely (\cite{GEOE97}).  Also, recent work (e.g., \cite{N97})
suggests that the Galactic Center wind is dominated by a few hot stars (in
particular, IRS 13E1 and 7W) with wind velocites
of $\sim 1000 \kms$.  Although these differences in mass and wind velocity
should not qualitatively change our results, they will be addressed in future work.

A total of 90$^3$ active zones are used in the large scale simulation.  
The zones are geometrically scaled so that the central zones are 1/64th 
the size of the outermost
zones.   An inner boundary, one that can be thought of as a totally absorbing
sphere $0.1 R_A$ or $2\times10^{-3}\pc$ in radius, is used to simulate 
the presence of the central black hole; at every time step, the mass 
and internal energy density of zones in the sphere are set to small 
values, corresponding to a temperature of $10^4$ K.
Once equilibrium is reached, the 
hydrodynamical values of zones just outside the inner
boundary are saved every ten years for 100 years, to be used as time 
dependent outer boundary conditions for a
smaller scale simulation.  The large scale 
simulations show that most of the wind
that reaches the central $0.1 R_A$ is post-shock gas 
flowing in the -z direction.  Figure 1 shows the radial mass accretion
flux through this inner boundary versus the accretion polar
angle, $\theta$, where $\theta=0$ corresponds to accretion along 
the +z axis, for a particular moment in time; 
the shape of the curve is not very time-dependent.  The clear
peak at $\theta \sim 135^{\circ}$ illustrates that back-flow is dominant.
That is, the flow forms a standing bowshock (or, more accurately, a compressional
wave) in front of the accretor and a low density, wide-angle ``Bondi tube'' behind it.
Figure 2 shows the total mass accretion rate through
the large scale inner boundary for the 100 years that are used for
the small scale simulations.  The average mass accretion rate is
$8.9\pm0.3\times{10}^{-5}\;M_\odot$ yr$^{-1}$, in good agreement with
the estimate of $9.3\times{10}^{-5}\;M_\odot$ yr$^{-1}$ from Equation 1.

\begin{figure}\label{fig-vstheta}
\epsscale{0.40}
\plotone{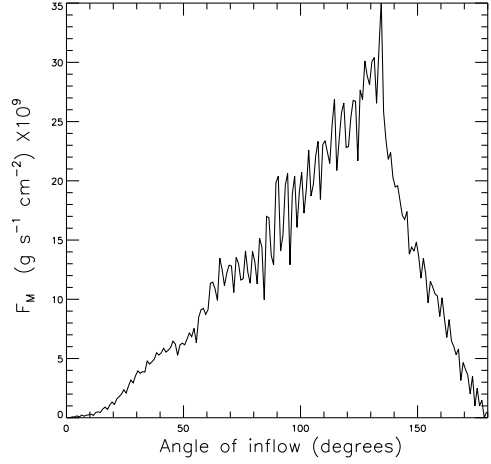}
\caption{A plot of radial mass flux in $\gscm2$ versus inflow 
polar angle $\theta$, summed into 1 degree bins.}
\end{figure}
\begin{figure}\label{fig-bigmdot}
\epsscale{0.40}
\plotone{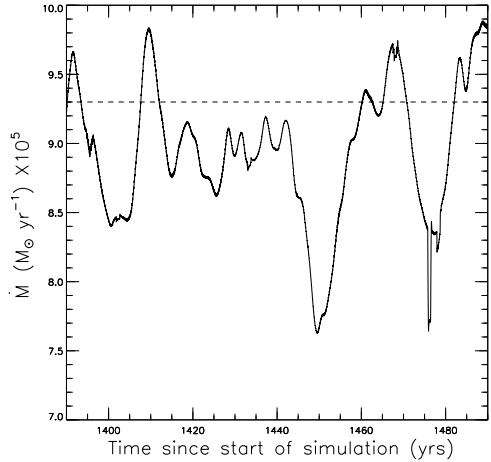}
\caption{A plot of mass accretion rate (in solar masses 
per year) versus time since the
start of the simulation (in years) for the 100 year 
timeslice of the large scale run that is
used in the small scale simulations.  Data are moving-box 
averaged over 40 time steps ($\sim 0.5$ year).
The dashed line is the theoretical value from Equation 1.}
\end{figure}

The small scale simulations, with 70$^3$ zones and a 
zone size scaling ratio of 1.01 (so that the outermost zones are roughly
1.4 times the size of the central zones), are 
1/15th of an $R_A$ or $1\times10^{-3}\pc$ or $1.2\times10^4\,R_s$ on a side.
The accretion disk is simulated as a totally 
absorbing disk 1/50th of an $R_A$ in radius and
3 zones ($\sim 300 R_s$) thick.  One simulation has the 
normal to the surface of the disk
pointing along the z axis (the ``parallel'' case) while one has the 
normal along the y axis (the
``perpendicular'' case).  The large scale inner boundary results are 
linearly interpolated in
time and trilinearly interpolated in space to 
produce outer boundary conditions, updated
every time step, for the small scale simulations.  If the large scale
results dictate that a given small scale outer boundary
zone does not have gas flowing into the volume of solution, that zone's
boundary condition is set to outflow; during our simulations,
this outflow has amounted to $<20\%$ of the
total mass inflow, which therefore represents an uncertainty of
this magnitude in the final results.
Note that the angular momentum in this outflow is minimal, since, as 
seen in Figure 1,
the mass accretion flux approaches zero only along the z axis.

\subsection{Results of the Hydrodynamical Simulations}
The intent of the numerical simulations is to more
realistically estimate the radial and azimuthal distribution of 
the infalling gas.  The characteristics of the gaseous distribution,
including the density contours and velocity vector fields
at the end of the two small scale simulations, are shown
in Figures 3-6.  

\begin{figure}\label{fig-rhopara}
\epsscale{0.40}
\plotone{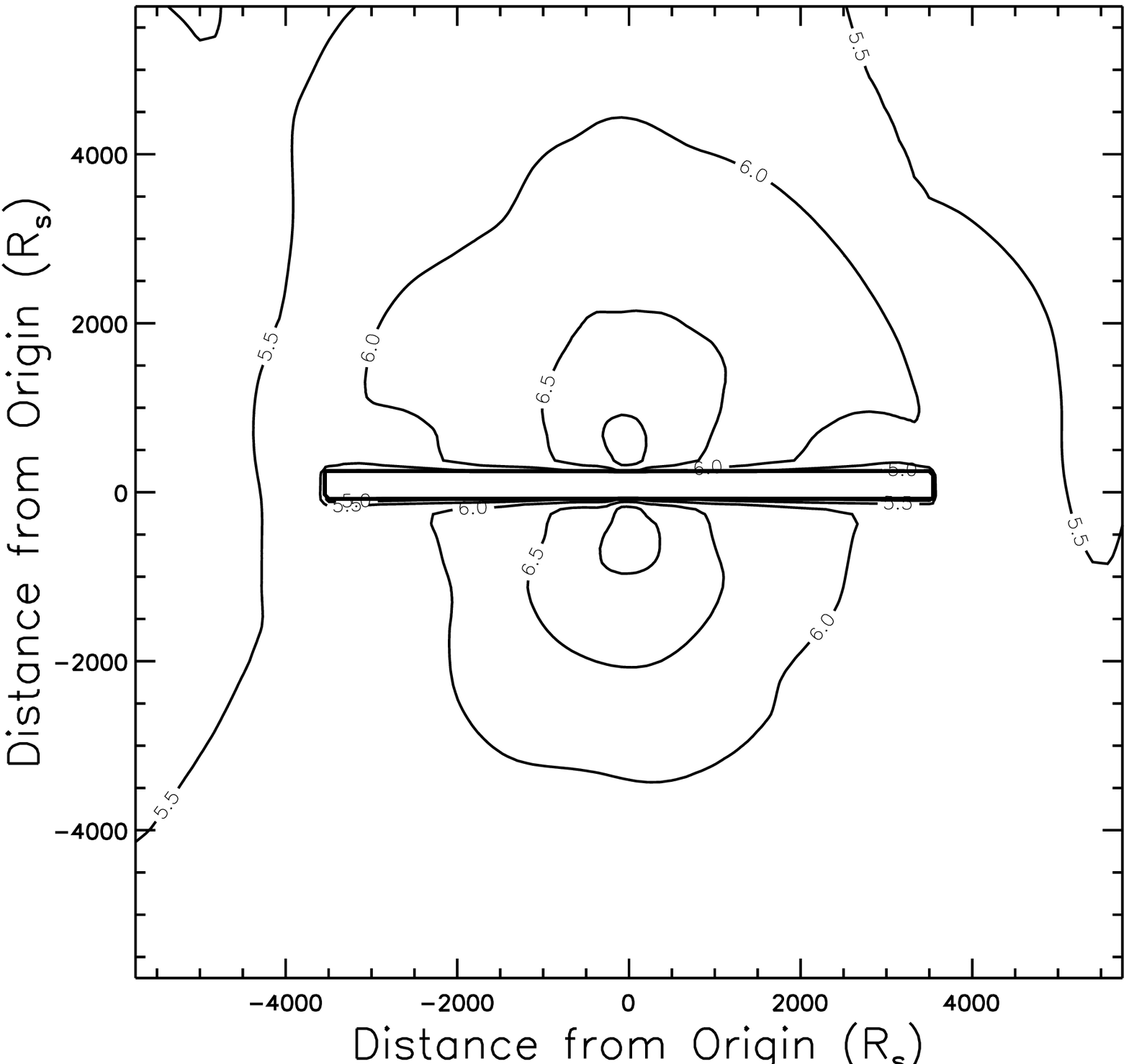}
\caption{Logarithmic density contours of a slice through x=0 for the 
small scale parallel simulation.  The large scale uniform wind 
originally comes from above.
There are 15 contours, ranging from $10$ to ${10}^{7} \ncm3$.
Note that the Schwarzschild radius ($R_s$) for a ${10}^6\msun$ black 
hole is $3\times{10}^{11}\cm$.}
\end{figure}
\begin{figure}\label{fig-rhoperp}
\epsscale{0.40}
\plotone{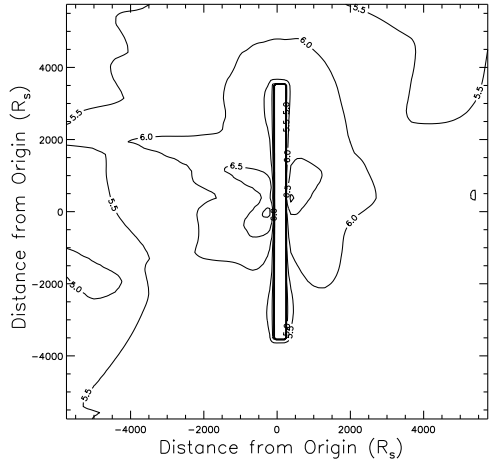}
\caption{Same as Figure 3 but for the perpendicular simulation.}
\end{figure}
\begin{figure}\label{fig-velpara}
\epsscale{0.40}
\plotone{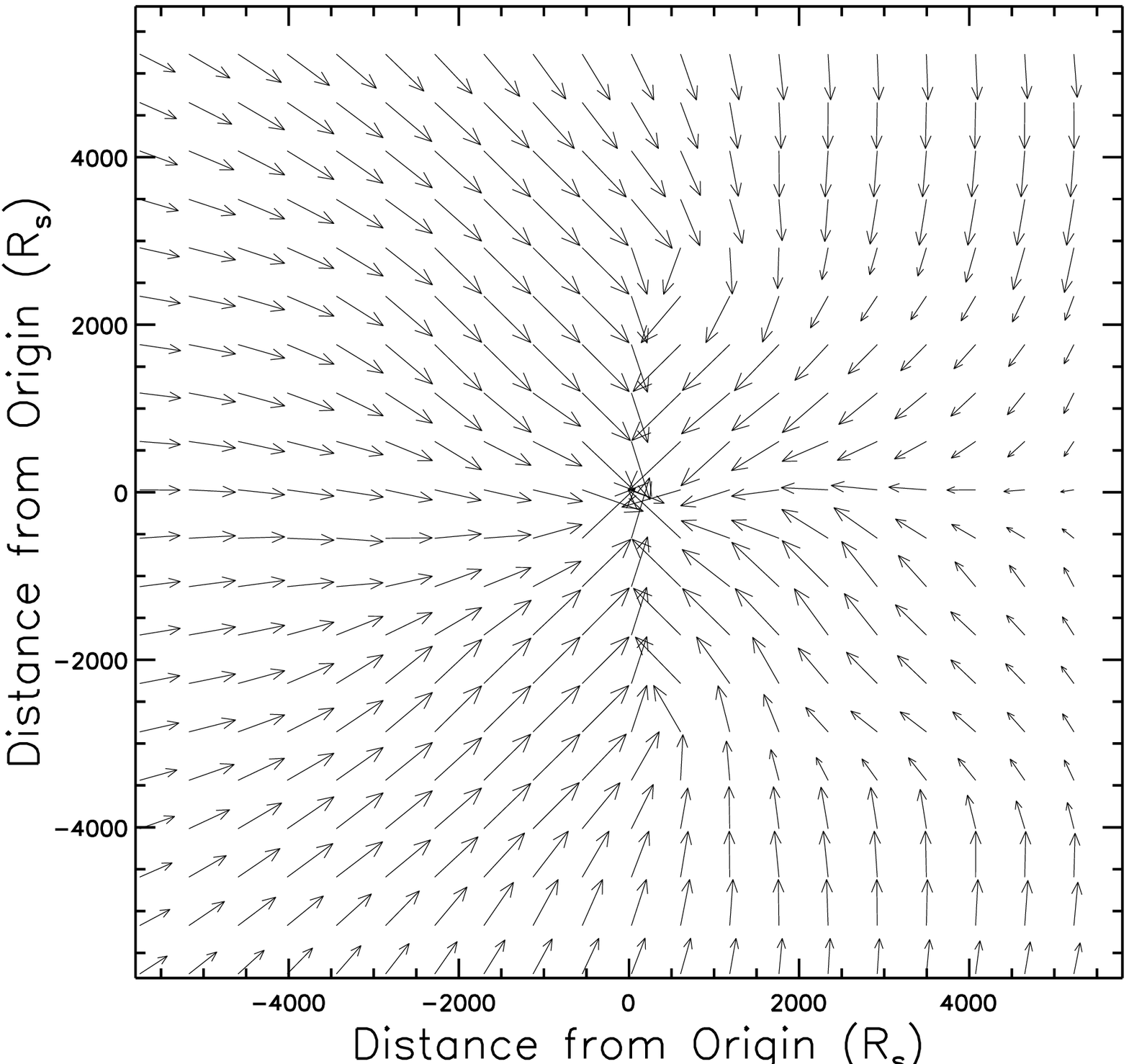}
\caption{The 2 dimensional velocity vector field through x=0 for the 
small scale parallel simulation with the same orientation as Figure 3. 
Note that most of the gas is flowing in from below.  The 
vectors are linear with the largest one
corresponding to a velocity (in the x=0 plane) of $\sim 3100 \kms$.}
\end{figure}
\begin{figure}\label{fig-velperp}
\epsscale{0.40}
\plotone{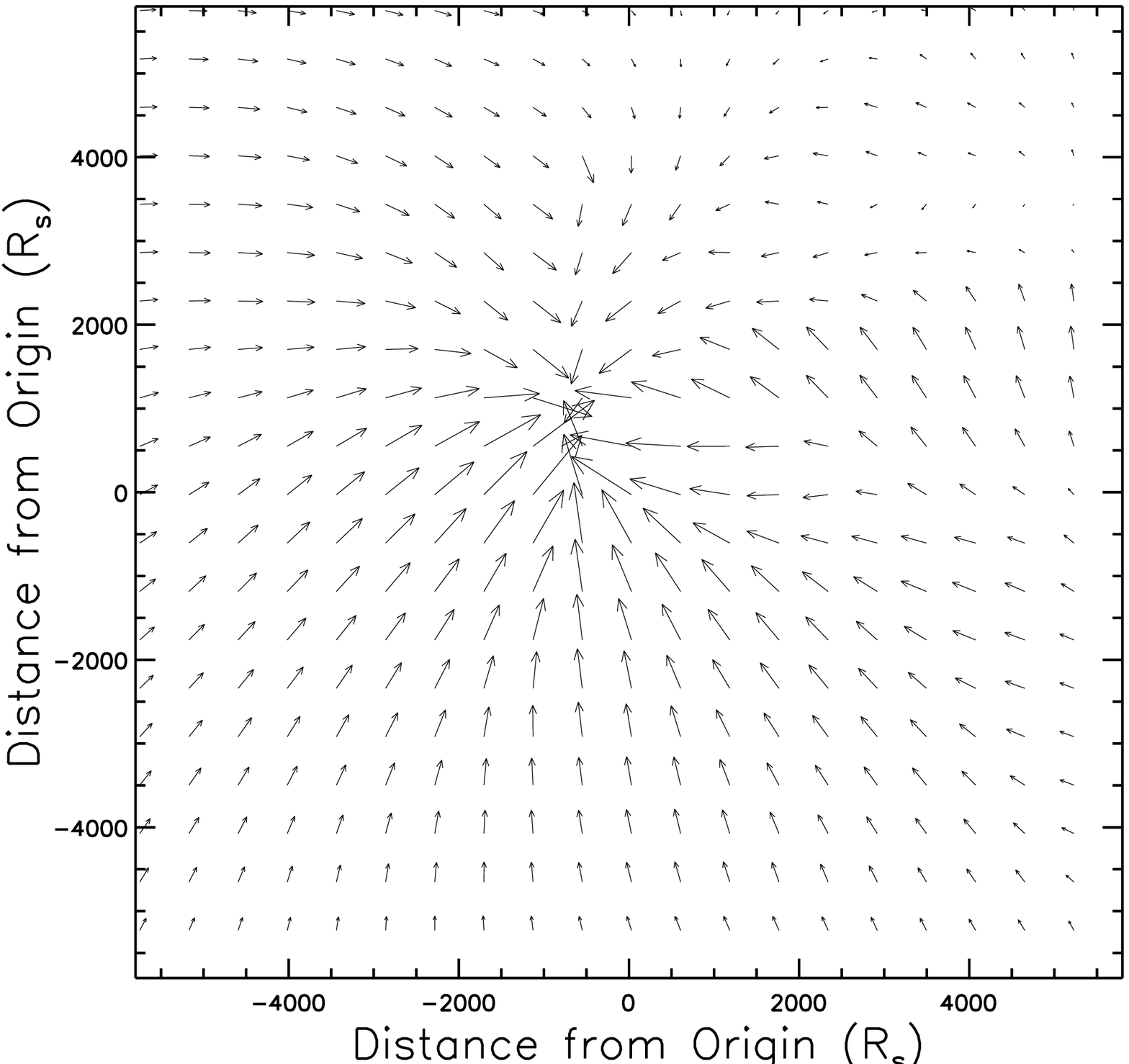}
\caption{Same as Figure 5 but for the perpendicular simulation.  The longest
vector corresponds to a velocity (in the x=0 plane) of $\sim 8800 \kms$.}
\end{figure}

In Paper I, it was assumed that the temperature of the infalling
gas was small compared to the resulting shock temperature, $T_{sh}$,
when the wind's kinetic energy is converted into thermal energy upon
impacting the disk.  That is, the gas was assumed to be highly supersonic.
Figure 7 shows the averaged Mach number for the infalling wind as it
impacts the disk.  On average, the gas is indeed supersonic, but, particularly
at moderate radii for the parallel run, individual zones reach Mach numbers
as low as $\sim 1.1$.  Due to the presence of numerical viscosity,
the Mach number and the other hydrodynamical
variables are variable both spatially and
temporally so Figure 7 (and similar figures shown below)
are only illustrative.  In Figure
8 we show the temperature of the infalling wind versus radius as well as 
the temperature of the disk that results from converting the wind's relative
kinetic energy into heat (according to Eq. 3 in Paper I 
with a correction factor of 1/18).
The thick solid line corresponds to the perpendicular run while the thin one is
from the parallel run (see Figure 9 below).  Clearly, the assumption that
$T_{sh}$ is much larger than $T_w$, the temperature of the infalling wind, is 
generally but not always valid.  The approximation is particularly poor for the
parallel run.  Note, however, that
the hydrodynamical code does not include radiative cooling or relativistic
corrections.
\begin{figure}\label{fig-machvsr}
\epsscale{0.40}
\plotone{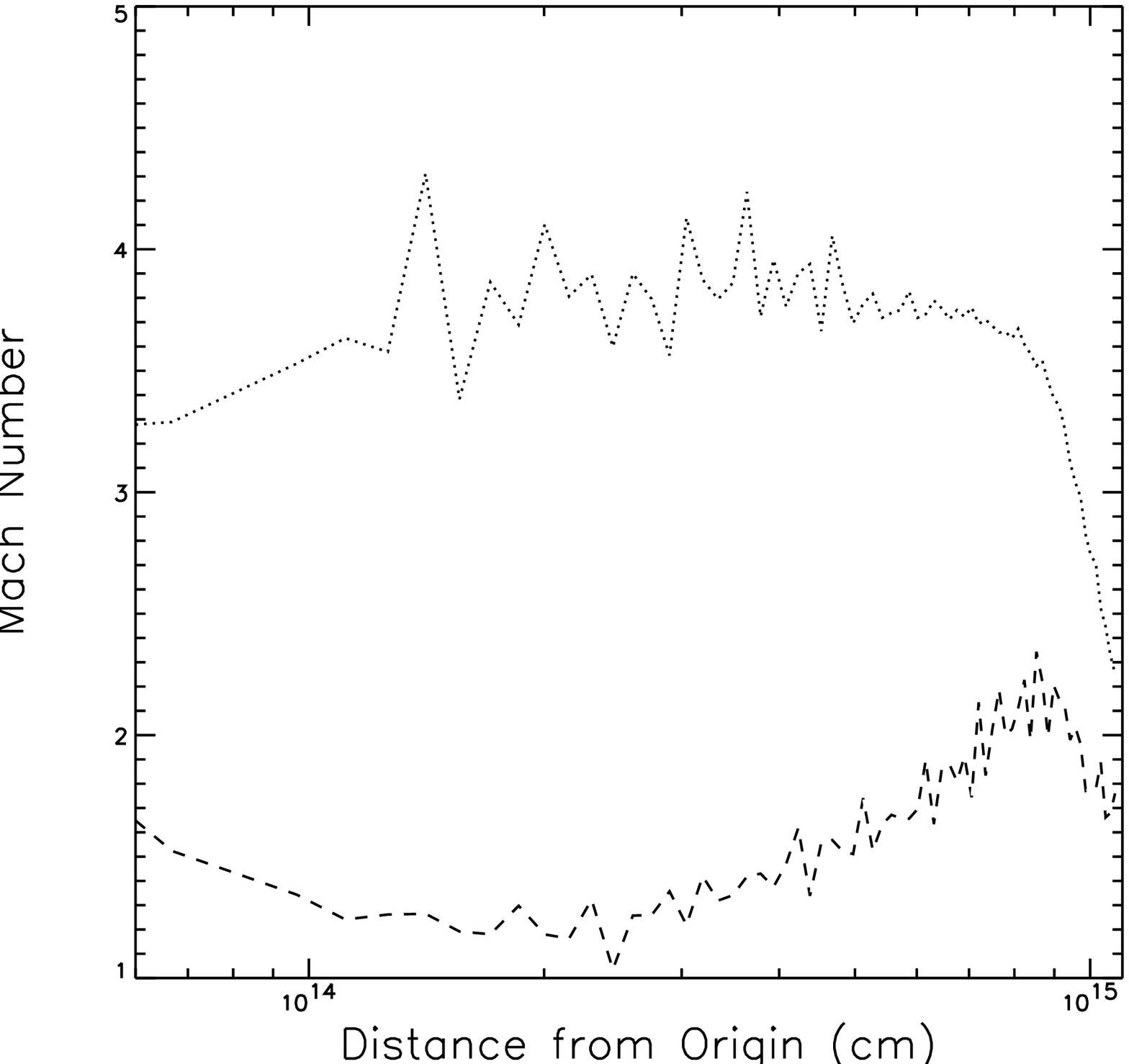}
\caption{Plot of the Mach number, averaged over $\phi$ and the two
sides of the disk, of the gas impacting the fossil disk at t = 50 years.
The dotted line corresponds to the
perpendicular disk simulation; the dashed line corresponds to the
parallel disk simulation.}
\end{figure}
\begin{figure}\label{fig-Tvsr}
\epsscale{0.40}
\plotone{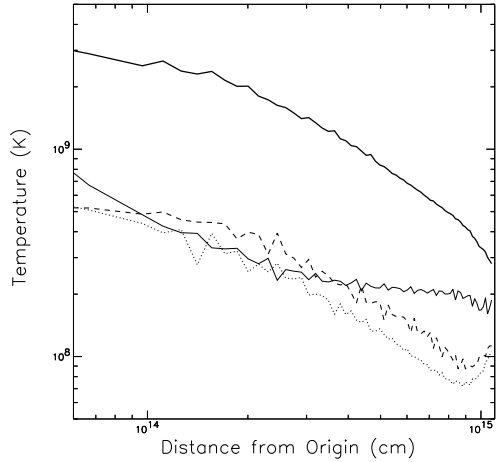}
\caption{Plot of the temperature of the infalling gas right
before it impacts the disk.  The dotted line corresponds to the
perpendicular disk simulation; the dashed line corresponds to the
parallel disk simulation.  The solid lines correspond to the resulting 
temperature for the two runs after the relative kinetic energy of the 
wind is converted into thermal energy. The thick solid curve is
for the perpendicular run and the thin one is for the parallel run.}
\end{figure}

However, the assumption that the wind is in free-fall from infinity is not
accurate, as Figure 9 illustrates.  The angular momentum in the wind and the obstructing
presence of the disk itself cause the wind to have a non-radial profile
and thus not fit the $r^{-1/2}$ free-fall expression used in Paper I.
The total mass accretion rate onto the disk for both runs is $\sim 4 \times 10^{21} \gms$;
this is somewhat less than the total mass flow into the volume ($5.6\times 10^{21} \gms$)
due to the outflow mentioned above.
A plot of the resulting $\dsigw$, shown in Figure 10, is slightly flatter
than a $r^{-2}$ dependence, suggesting, as expected from Figure 1,
a wind pattern that is somewhere between cylindrical and radial.  
That is, for a cylindrical flow, $\dsigw$ is constant with
radius (in the plane), while for a radial flow intercepted by 
a disk of thickness $2d\ll r$,
\begin{equation}
r^3\,\dsigw \propto d\approx \hbox{constant}\;, 
\end{equation}
since $\dsigw=v_z\;\rho$, with $v_z=v_r\cos\theta$ and 
$\dot M=4\pi\,r^2\,\rho\,v_r$.

\begin{figure}\label{fig-velvsr}
\epsscale{0.40}
\plotone{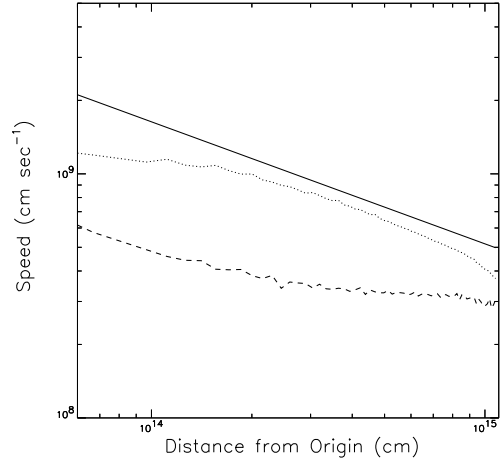}
\caption{As in Figure 7, but for the speed
of the gas impacting the fossil disk.  
The solid line corresponds to free-fall velocity from
infinity.}
\end{figure}
\begin{figure}\label{fig-sdotvsr}
\epsscale{0.40}
\plotone{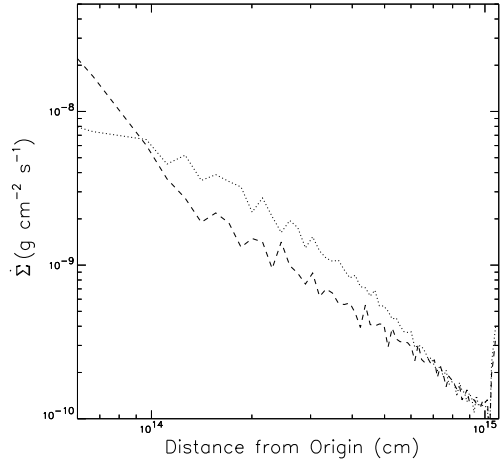}
\caption{As in Figure 7, but for $\dsigw$, the total mass infall onto the disk 
per unit time and area.}
\end{figure}

In Figures 11 and 12 we show the mass accretion rate onto the accretion
disk versus time for the parallel and perpendicular runs, respectively.
A comparison with Figure 2 shows that $10-20\%$ of the gas is not falling
onto the disk but rather is leaving the volume of the small scale simulation
through the +z boundary.  Also, note that the large scale mass accretion rate
is, on the whole, temporally uncorrelated with the 
small scale mass accretion rate
but that the accretion rate for the two runs are strongly correlated.  Thus,
the average rate of mass falling onto the disk is 
independent of the disk orientation while
the velocity profile, and thus the specific angular momentum, are not.
However, coincidentally, the angle between the normal of the bottom of the disk (90 or 180$^\circ$)
and the angle of maximum inflow (135$^\circ$, see Figure 1), is 45$^\circ$ for both
runs.  A disk oriented differently with respect to the large scale inflow might
result in different accretion profiles (but see section 4.3).

\begin{figure}\label{fig-mdotpara}
\epsscale{0.40}
\plotone{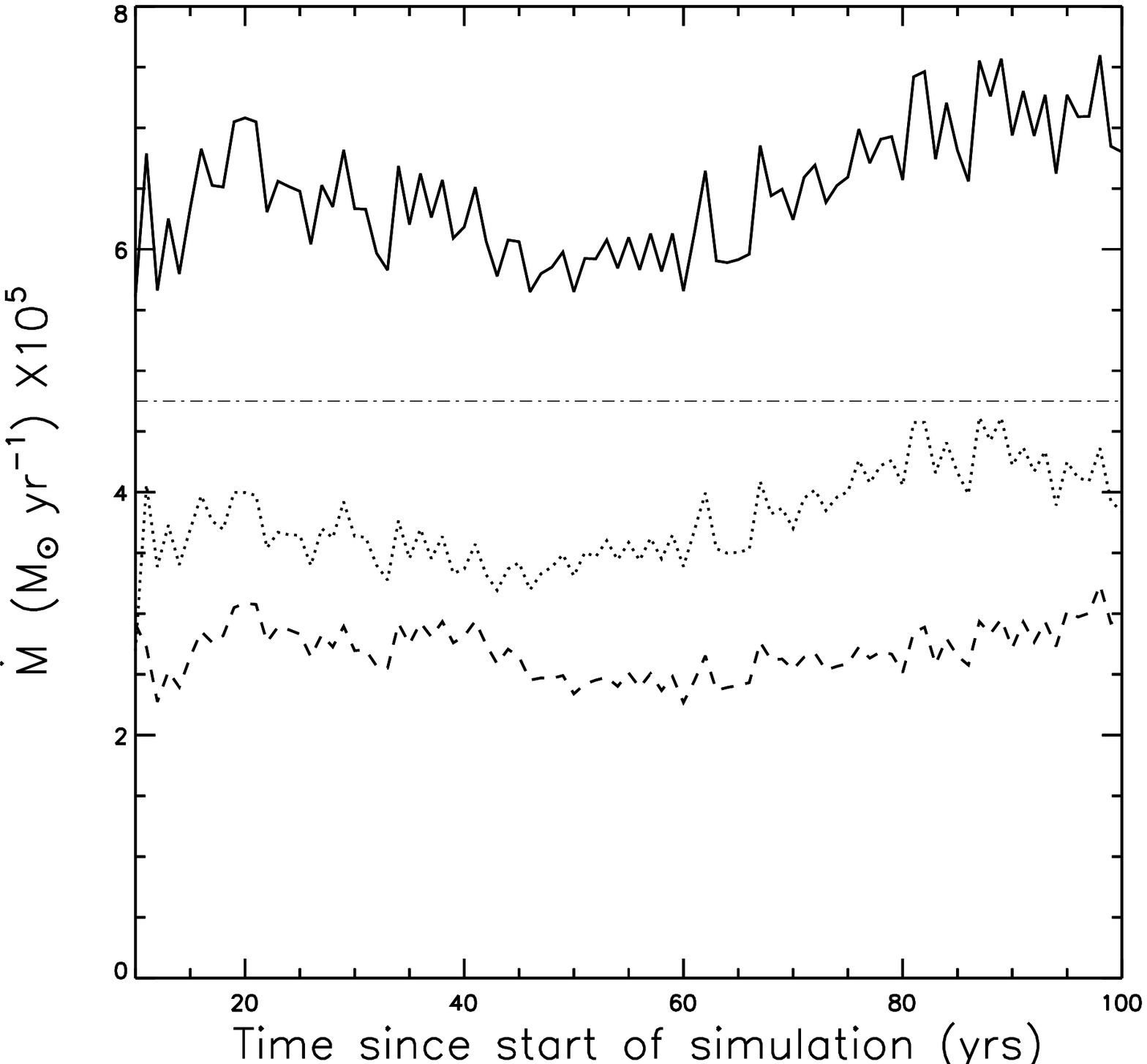}
\caption{The mass accretion rate versus time onto the accretion disk due
to the infalling wind for the parallel disk case.  The dotted line is for the 
``top'' side (+z) of the disk, the dashed line is for the ``bottom'' side,
and the solid line is the sum of the two.  The dot-dashed line corresponds
to one half of the theoretical value from Equation~\ref{mdot}, for
direct comparison with the top and bottom rates.}
\end{figure}
\begin{figure}\label{fig-mdotperp}
\epsscale{0.40}
\plotone{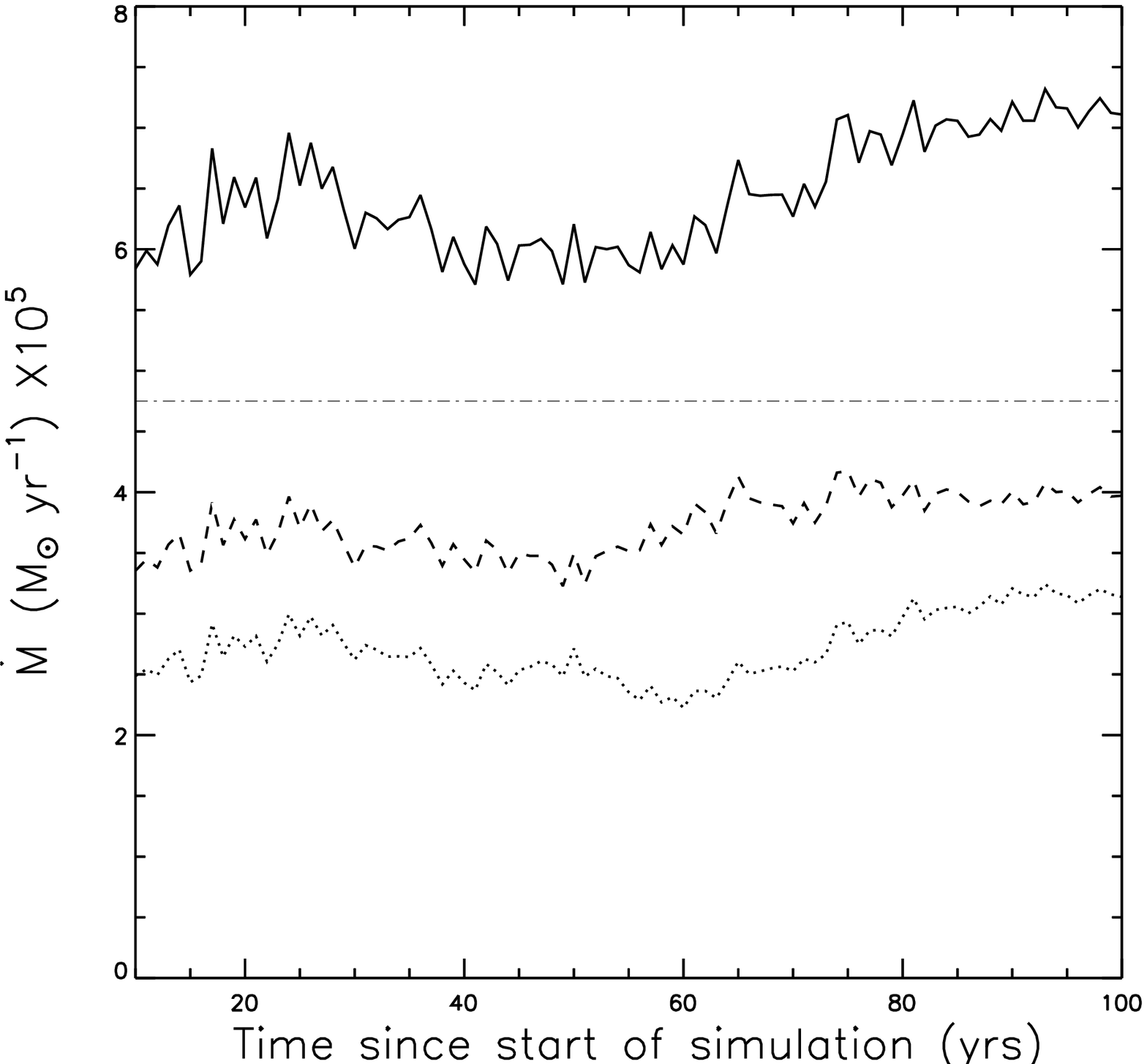}
\caption{As in Figure 10, but for the perpendicular disk case.}
\end{figure}

Similarly, Figures 13 and 14 show the specific angular momentum, $\lambda$,
in units of $c R_s$, accreted onto the accretion disk via the wind
for the parallel and perpendicular runs, respectively.  In contrast
to the mass accretion rate, the magnitude of $\lambda$ is significantly
different for the two runs, the parallel case having 
$\lambda = 1.6\pm0.2$, which is roughly 4 times lower than 
for the perpendicular case with $\lambda = 6.8\pm4.1$.
The flow in the perpendicular case is considerably more turbulent, with
larger temporal fluctuations in the hydrodynamical variables.
It is expected that a disk with a normal along 
the large scale z axis would accrete less specific angular momentum
since $L_z$ is, in general, less than the other components, $L_x$ and $L_y$.
The two disk orientations can be thought to 
represent extremes of minimal and maximal capture of the
specific angular momentum in the infalling wind.  The effective circularization
radius, where most of the wind intercepts the disk, is $\sim 4 R_s$ for the
parallel case and $\sim 56 R_s$ for the perpendicular case.  It is important to note
that the specific angular momentum accreted by the disk is always less than
the specific angular momentum present in the overall wind.  Thus, even though
recent large scale simulations (\cite{CM97}), involving point sources 
instead of a planar flow, result in
larger values of $\lambda$ ($\sim 50$ at $5\times10^{15}$ cm), any accretion disk 
is likely to intercept
only a fraction of this because clumps with preferentially small
values of $\lambda$ tend to flow inwards to smaller radii.

\begin{figure}\label{fig-lampara}
\epsscale{0.40}
\plotone{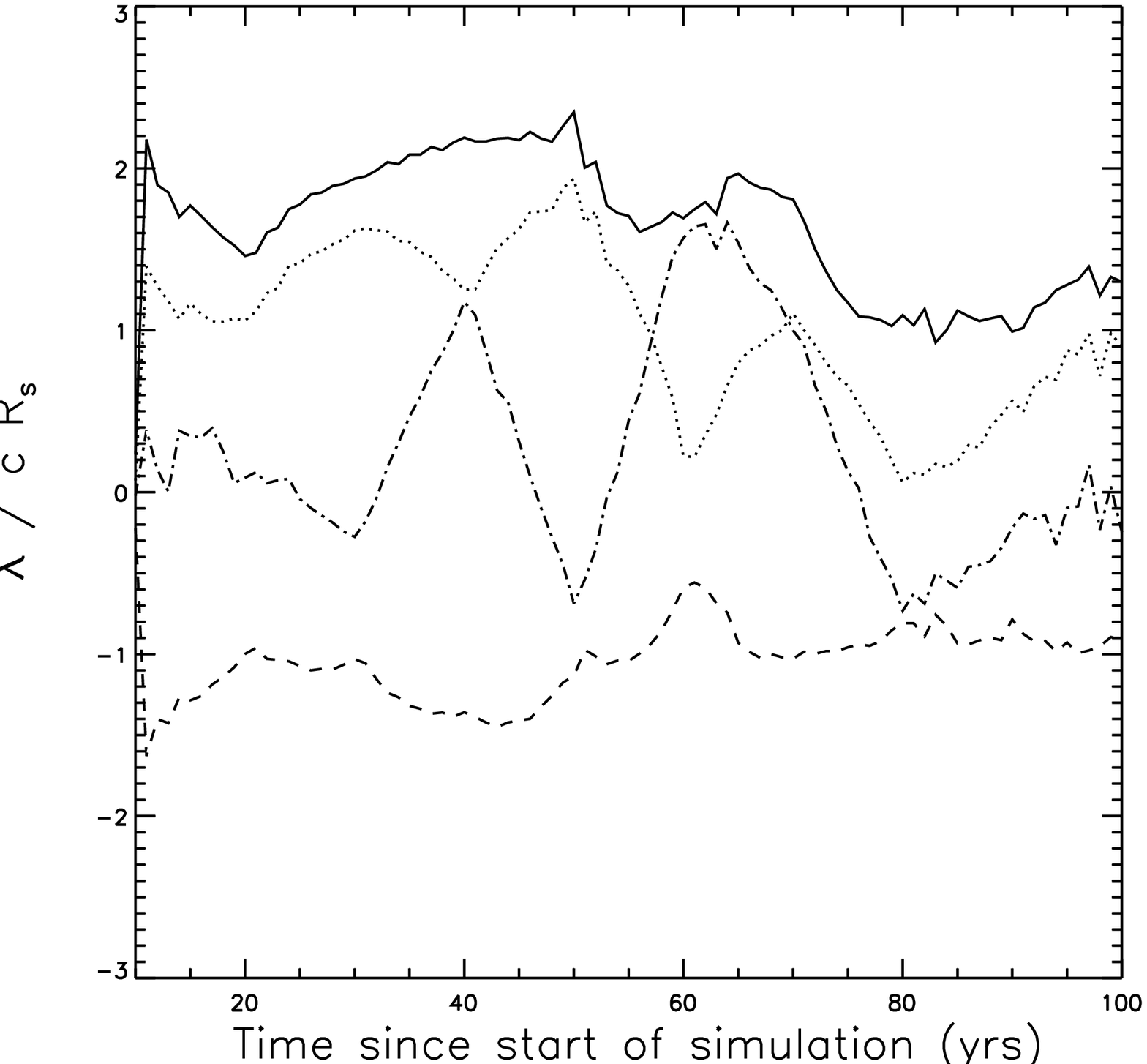}
\caption{The specific angular momentum accreted by the disk versus time
for the parallel case.
The dotted line is $\lambda_x$, the dashed line is $\lambda_y$, and the
dot-dashed line is $\lambda_z$. 
The solid line is the root-sum-square of the components.  }
\end{figure}
\begin{figure}\label{fig-lamperp}
\epsscale{0.40}
\plotone{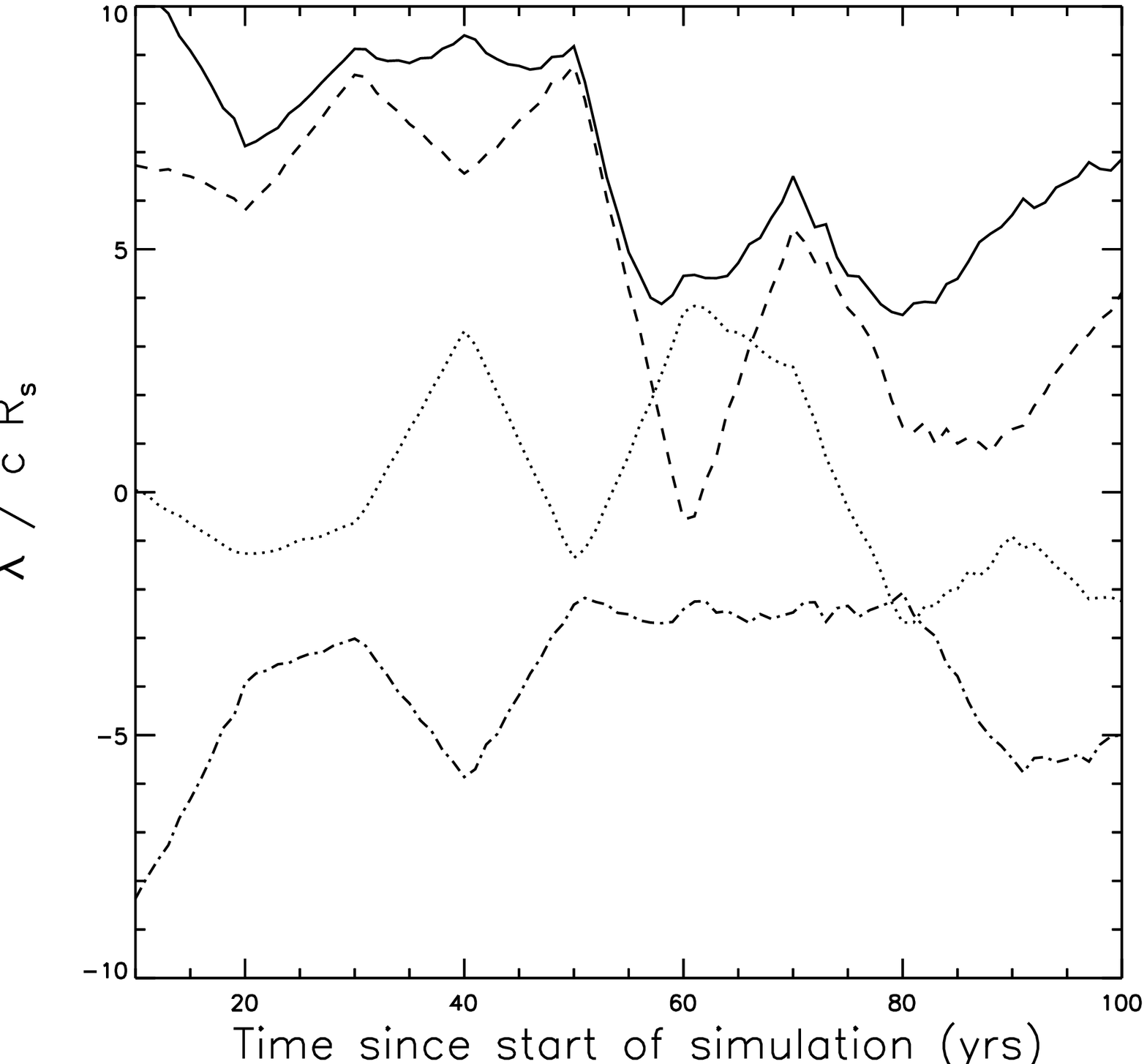}
\caption{As in Figure 12, but for the perpendicular disk case.  Note that for
this plot, $\hat z$ is parallel to the normal to the disk surface, corresponding
to the large scale y axis.}
\end{figure}

In Figures 15-18, we present some of the velocity profiles, in cylindrical coordinates, for the
wind as it falls onto the disk.  For comparison, the radial velocity for
a shocked gas in free-fall, corresponding to the analytical profile used in
Paper I for the zero angular momentum case, is also plotted in the figures.
Consistent with the small value of $\lambda$, the velocity
profile for the parallel case is fairly close to the shocked
free-fall profile used in Paper I while $v_r$ for the perpendicular
case is, except at small $r$, somewhat {\sl faster} than the profile
used in Paper I.  This is
a direct consequence of the non-zero velocity at infinity.  In addition,
the large scale bowshock is not a true shock for all angles so that a shocked
free-fall profile, which is 1/4 times that of an unshocked profile, is inaccurate. 
\begin{figure}\label{fig-v50tpara}
\epsscale{0.40}
\plotone{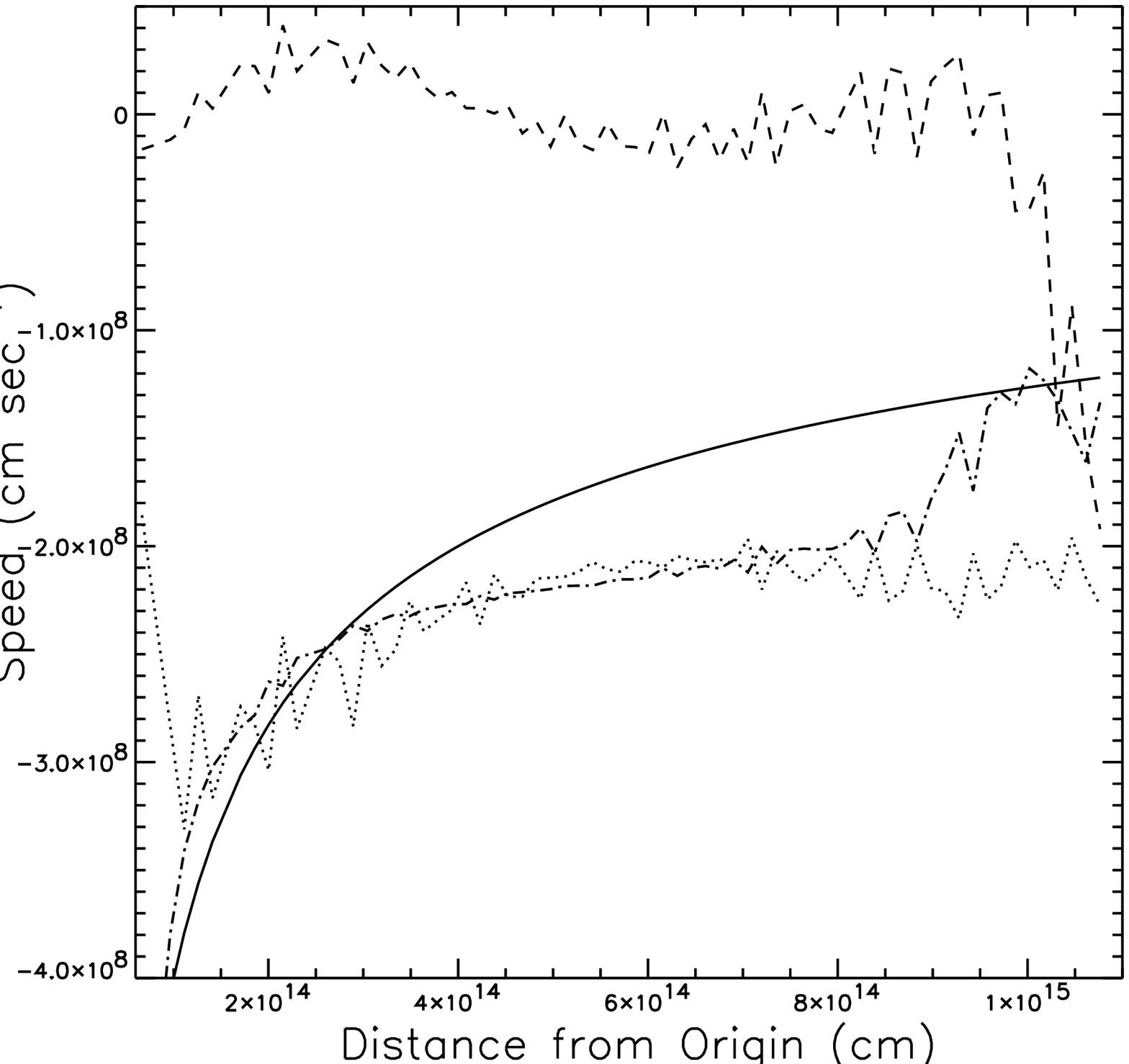}
\caption{The velocity profile for gas falling onto the ``top'' side of the disk
for the parallel run at t = 50 years.
The dotted line is $v_r$, the dashed line is 
$v_{\phi}$, and the
dot-dashed line is $v_z$.
The solid line is the $r^{-1/2}$
profile of a shocked free-falling gas (which is 1/4 the velocity of a gas
free-falling from infinity).}
\end{figure}
\begin{figure}\label{fig-v50bpara}
\epsscale{0.40}
\plotone{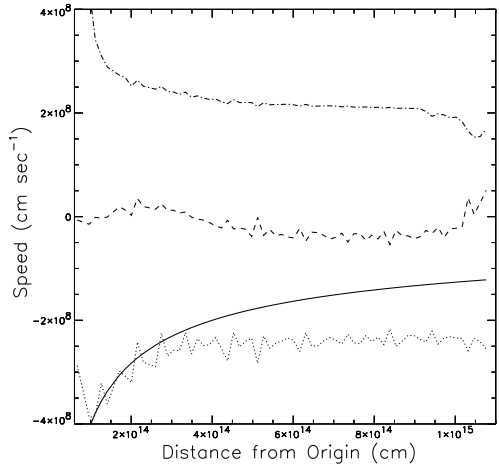}
\caption{As in Figure 15, but for the ``bottom'' side of the disk.}
\end{figure}
\begin{figure}\label{fig-v50tperp}
\epsscale{0.40}
\plotone{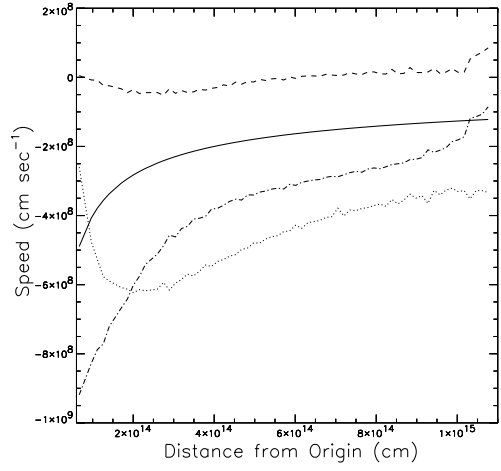}
\caption{As in Figure 15, but for the perpendicular run.  Note that for
this plot, $\hat z$ is parallel to the normal to the disk surface.}
\end{figure}
\begin{figure}\label{fig-v50bperp}
\epsscale{0.40}
\plotone{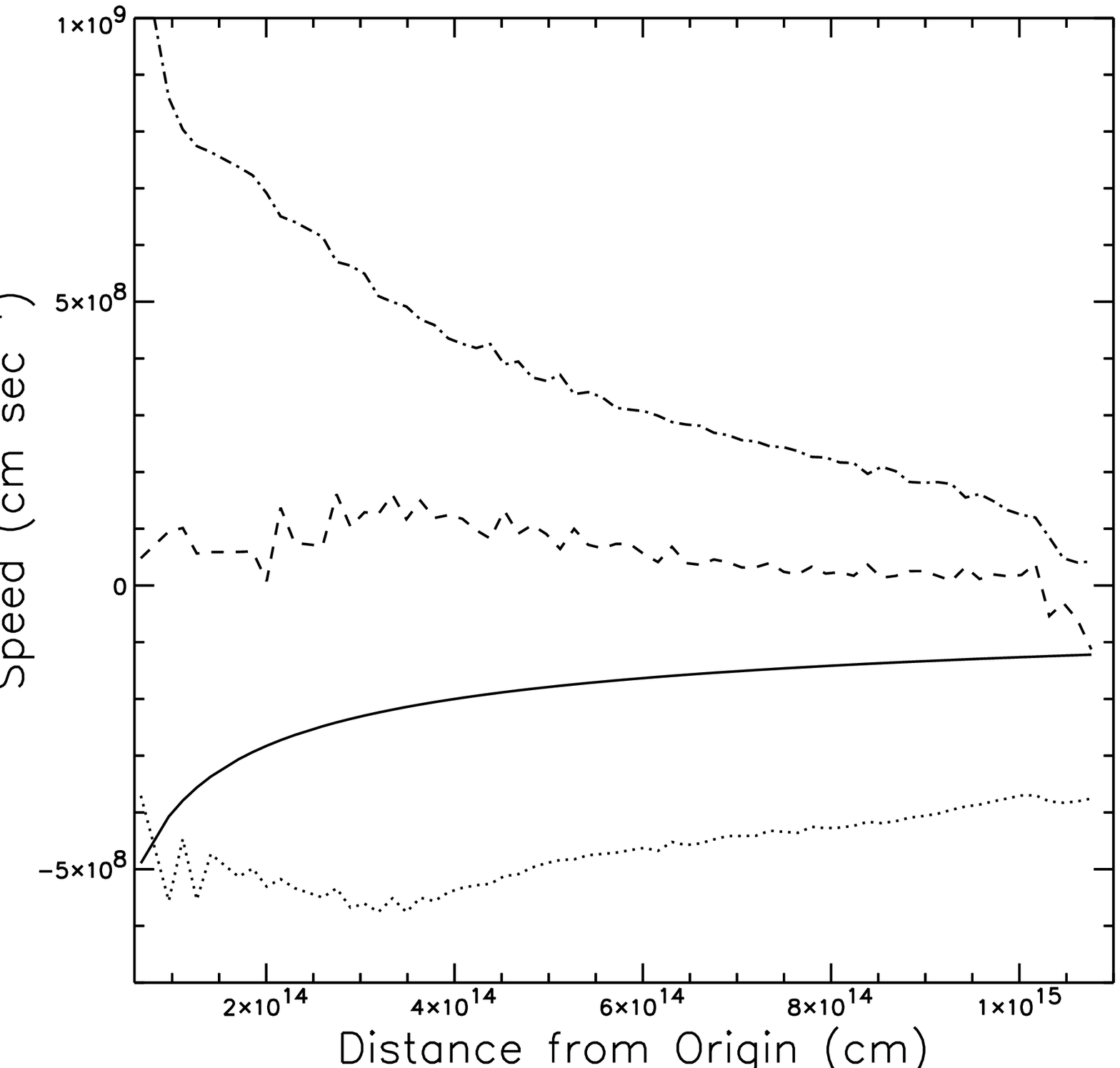}
\caption{As in Figure 17 but for the ``bottom'' side of the disk.  For
this plot, $\hat z$ is parallel to the normal to the disk surface.}
\end{figure}
Figure 19 illustrates quantitatively how the specific angular momentum 
present in the gas accreting onto the disk relates to that of the disk itself, which is
assumed to be Keplerian so that $\lambda_d = \sqrt{GMR}$.
\begin{figure}\label{fig-lkepvsr}
\epsscale{0.40}
\plotone{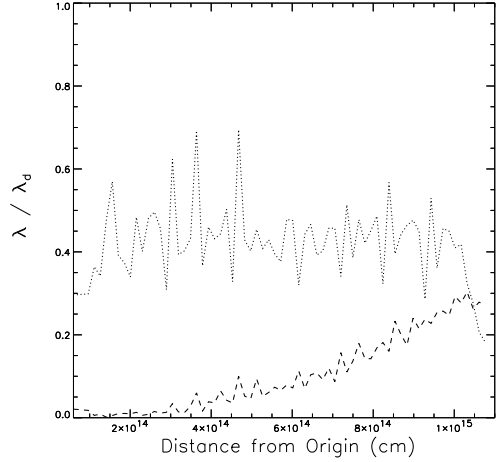}
\caption{Plot of $\lambda$/$\lambda_d$, averaged over $\phi$ and the two
sides of the disk, at t = 50 years.  The dotted line corresponds to the
perpendicular disk simulation; the dashed line corresponds to the
parallel disk simulation.}
\end{figure}

For the purpose of using the 3D hydrodynamical results for the quasi-analytical 
disk evolution calculations below, we have averaged the flow variables over 
$\phi$.  However, due to the asymmetric and time-dependent outer boundary conditions, the
flow has azimuthal structure. Figures 20-23 show where the specific angular momentum 
is deposited on the disk.  The lack of symmetry is particularly evident near the edge of the
disk, where its absorbing boundary induces instabilities in
the inflow.  More specific angular momentum is deposited in the central region for the
perpendicular calculation, consistent with the larger average value of $\lambda$
in this case.  Although the magnitude of the fluctuations in
$\lambda$, both spatially and temporally, are sometimes
large (for example, note the difference between Figures 22 and 23), 
the time scale for these variations is short compared with the
disk evolution time scale (see below) and these variations are therefore
not likely to affect our conclusions.
\begin{figure}\label{fig-l50tpara}
\epsscale{0.40}
\plotone{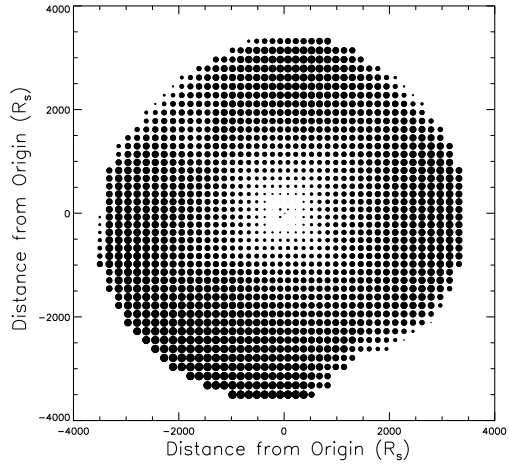}
\caption{A graphical representation of the specific angular momentum deposited 
on the top side of the disk by the infalling wind for the parallel calculation.  
The diameter of the black circles is proportional to the logarithmic value
of $\lambda$; circles that are just touching correspond to $\lambda \approx 10$.
}
\end{figure}
\begin{figure}\label{fig-l50bpara}
\epsscale{0.40}
\plotone{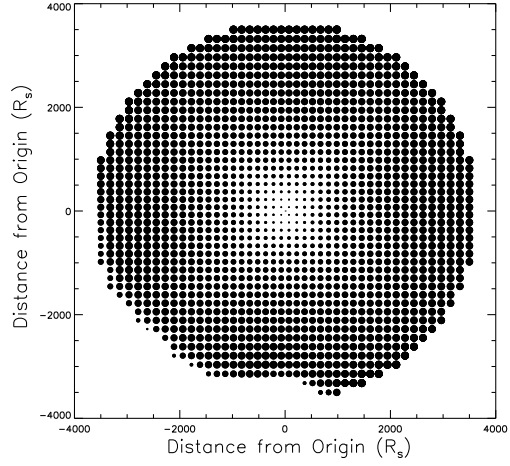}
\caption{As in Figure 20 but for the specific angular momentum deposited
on the bottom side of the disk for the parallel calculation.
}
\end{figure}
\begin{figure}\label{fig-l50tperp}
\epsscale{0.40}
\plotone{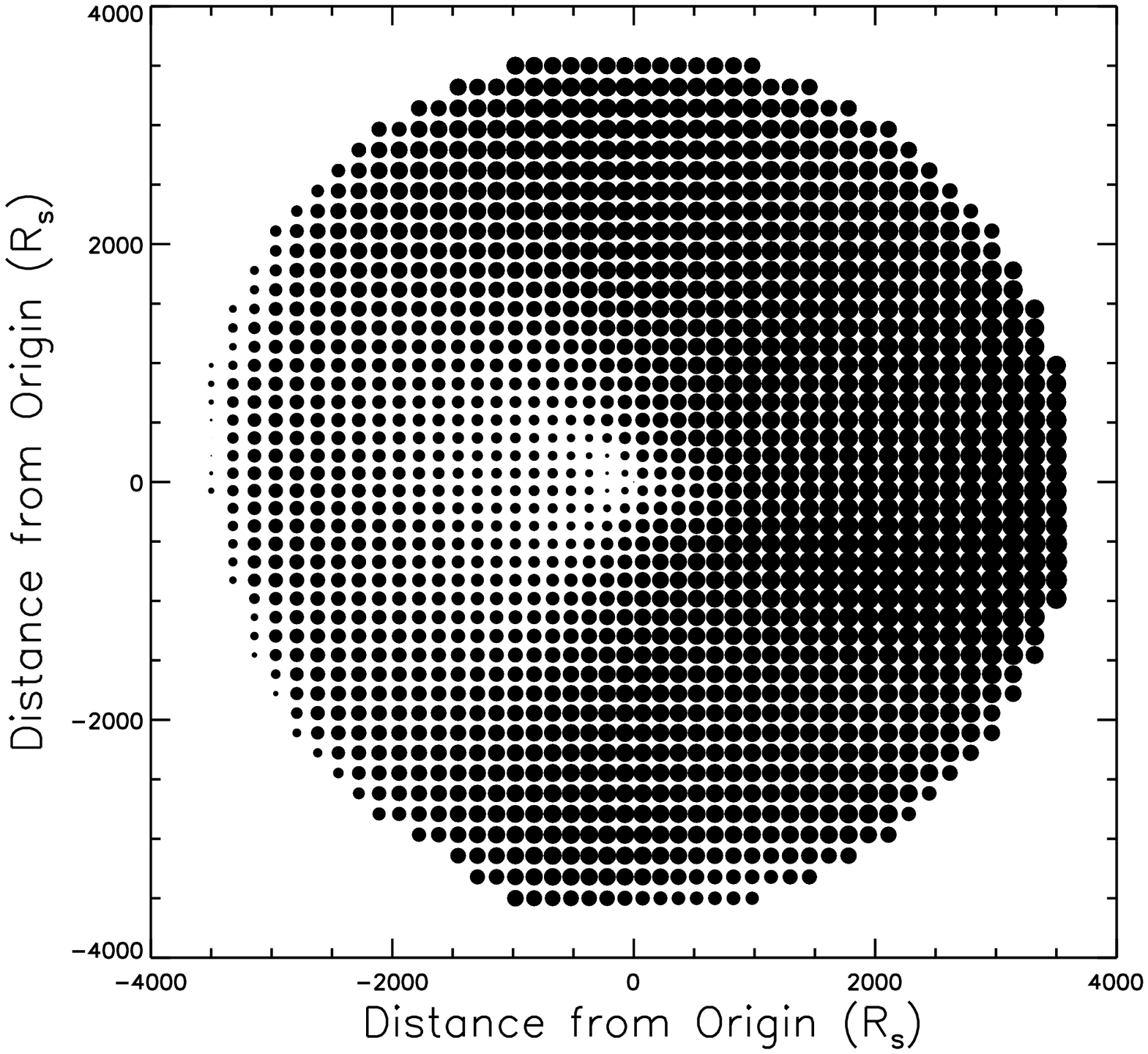}
\caption{As in Figure 20 but for the specific angular momentum deposited
on the top side of the disk for the perpendicular calculation.
}
\end{figure}
\begin{figure}\label{fig-l50bperp}
\epsscale{0.40}
\plotone{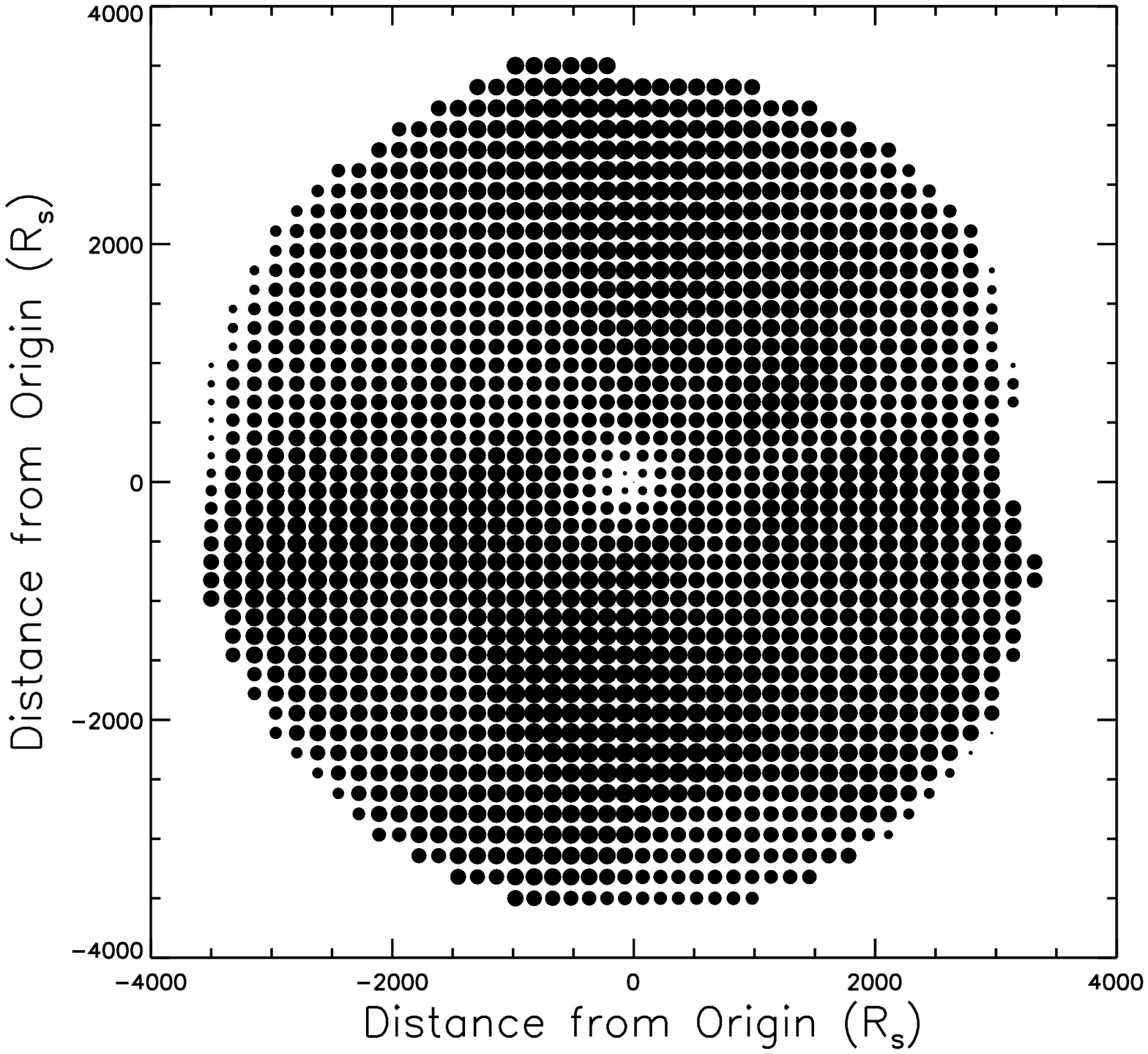}
\caption{As in Figure 20 but for the specific angular momentum deposited
on the bottom side of the disk for the perpendicular calculation.
}
\end{figure}

\subsection{Power-law Fitting}

As evidenced by the preceeding plots, a power-law fit to the velocity
profiles is not always accurate.  However, for our purposes, a simple
power-law allows for easy evolution of the disk equations.  In any case,
our final results are not overly sensitive to the precise exponents for
\dsigw\, $v_{\rm r}$, and $v_{\rm z}$.

The 3D hydrodynamical data are first averaged over $\phi$ and then put
into radial bins of 50 R$_s$.  This bin size was chosen so as to give
some idea of the magnitude of the fluctuations of the hydrodynamical
variables while still showing the large scale trends.
The volume of each zone was used as a
weight when averaging.  A given zone was used in calculating results for
the top or bottom side of the disk, depending on the sign of $v_z$.
These values were used in making the previous plots.

Next, a linear least squares fit was applied to the binned and averaged
data from years 50-99.  The results of these fits were then averaged to
get our final power-laws for \dsigw\, $v_{\rm r}$, and $v_{\rm z}$.  For
the perpendicular case we get:
\begin{eqnarray}
\dsigw(r)&=&1.35\cdot10^{17}\;{\rm g/cm^2}\;(r/{\rm cm})^{-1.8}\\
v_{\rm r}(r)&=&2.19\cdot10^{12}\; {\rm cm/sec}\;(r/{\rm cm})^{-0.25}\\ 
v_{\rm z}(r)&=&8.51\cdot10^{18}\; {\rm cm/sec}\;(r/{\rm cm})^{-0.71}\\ 
R_{circ}&=&56 R_s\;,
\end{eqnarray}
while for the parallel case we get:
\begin{eqnarray}
\dsigw(r)&=&1.17\cdot10^{11}\;{\rm g/cm^2}\;(r/{\rm cm})^{-1.39}+
1.07\cdot10^{17}\;{\rm g/cm^2}\;(r/{\rm cm})^{-1.8}\\
v_{\rm r}(r)&=&1.35\cdot10^{10}\;{\rm cm/sec}\;(r/{\rm cm})^{-0.12}\\
v_{\rm z}(r)&=&1.62\cdot10^{13}\;{\rm cm/sec}\;(r/{\rm cm})^{-0.33}\\
R_{circ}&=&4 R_s\;.
\end{eqnarray}
The velocity profile is the average of the results for the top and
bottom of the disk while \dsigw\ is the sum.
$R_{circ}$ is the circularization radius (given by $2 \lambda^2$)
determined from the time averaged value of $\lambda$.  As expected from
the small values of $\lambda$, the azimuthal velocity, $v_\phi$, is 
always smaller than $v_r$ and $v_z$ and is, on average, zero.
Note that Equation 20 has two components, one from each side of the
disk.  The flatter term arises from the bottom side of the disk.  This
is a result of the entire ``Bondi tube'' impacting the bottom of the disk
at relatively large radii;
in the perpendicular case, it impacts both sides of the disk.
This is also the reason for the flatter velocity profiles in the parallel case.

\section{The Disk Evolution with a Hydrodynamic Wind Infall}

We can now use the results of the 3D hydrodynamical calculations to
determine the temporal evolution of an accretion disk subject to a wind
infall. Here we will be using the 1D accretion disk code described in
Falcke \& Melia (1997). The code traces the evolution of a standard
accretion disk with infall according to the Equations of section 2,
which are employed with an arbitrary radial mass deposit as
input. A major problem in combining the 3D wind simulations with the
1D accretion disk calculations is that the length and time scales are
vastly different between the two. The 3D simulations do not have
sufficient resolution to model the infall at small radii.  Moreover,
they span only a period of a few hundred years. On the other hand,
the disk evolves on a scale of $10^{4-5}$ years. In addition,
the disk code can neither take into account azimuthal asymmetries of the
inflow, nor properly handle the differences in infall between the
top and bottom sides of the disk. One therefore has to make a number 
of simplifying assumptions.

We said earlier how one can roughly approximate the basic parameters
of the inflow as power-laws, e.g., for \dsigw\, $v_{\rm r}$, and $v_{\rm z}$. For 
\dsigw, the input for the disk simulations will then be the sum of 
top and bottom sides, while for $v_{\rm z}$ we take the average. The
internal energy of the wind in our 3D simulations is generally small compared
to its kinetic energy and will be neglected. We then have to assume
that these power-laws remain constant over the typical 
disk evolution time scale. We also need to assume that the power-laws can
be extrapolated down to the circularization radius. For
numerical reasons, we also truncate the wind infall at $10^{15}$cm,
and allow the actual accretion disk to extend outwards to three times
this distance.

The functional forms of \dsigw\ yield a wind infall rate of 
approximately $10^{-4} M_\odot$/yr. The inner wind radius is given by the
circularization radius.  The accretion disk evolution
always begins with the adoption of a Shakura-Sunyaev configuration
with $\alpha=10^{-4}$, $\dot M=10^{-6} M_\odot/$yr, and a central 
mass $M_\bullet=10^6M_\odot$, consistent with the 3D hydrodynamical
simulation. Following Falcke \& Melia (1997), we calculate the
disk's spectrum by integrating local blackbodies. We note that in the
parallel case the top and bottom sides receive different mass
depositions from the wind and hence are subjected to different
energy dissipation rates.  As such, the actual spectra of the top and 
bottom sides will generally not be the same.
Given our computational limitations and the fact that the modest
difference will not significantly change our conclusions, we are going
to ignore this effect in the following.

The results of our calculations are shown in Figures 24-27 for the 
perpendicular case and in Figures 28-31 for the parallel case.
Many of the results of our disk calculations that incorporate
the 3D hydrodynamical simulations can be understood in terms of the 
various possibilities outlined in our previous paper (Falcke \& Melia 
1997), but several interesting new features are now apparent, and
we discuss these in the next section.

\begin{figure}\label{sig2}
\epsscale{0.40}
\plotone{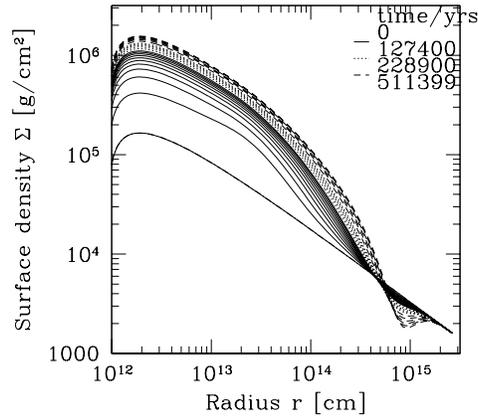}
\caption{Evolution of the surface density of an accretion disk with
wind infall when the incoming ambient flow is perpendicular to the disk's 
rotation axis. The disk is a standard
$\alpha$-disk with $\alpha=10^{-4}$ and $\dot M=10^{-6} M_\odot/$yr
around a black hole of $10^6 M_\odot$. Different lines correspond to
different time steps (equidistant for equal line styles) as shown in
the legend. The initial Shakura-Sunyaev solution coincides with the
first solid line.}
\end{figure}

\begin{figure}\label{mdot2}
\epsscale {0.40}
\plotone{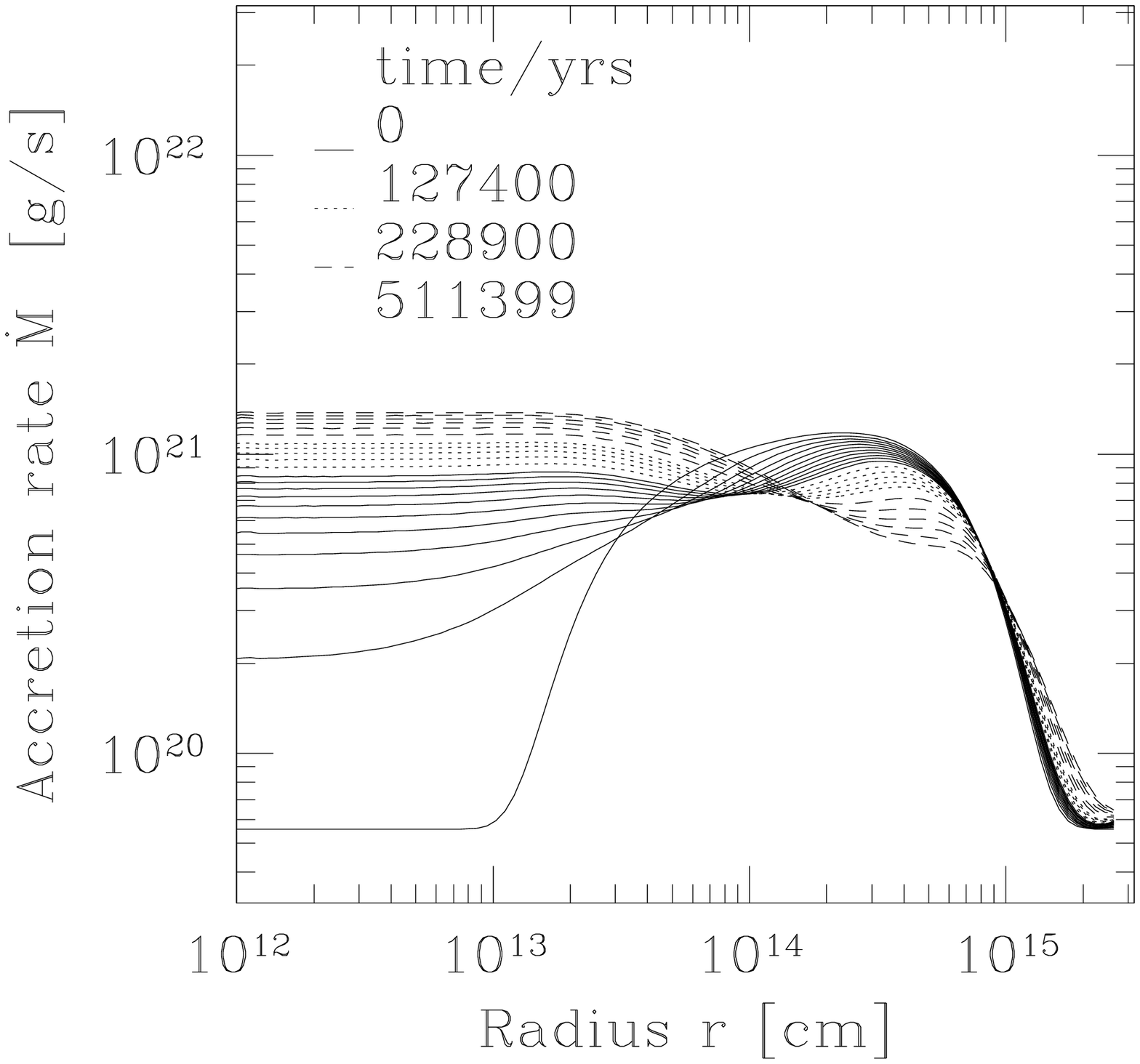}
\caption{Same as Figure 24, but here the local accretion rate in the disk
is shown.}
\end{figure}

\begin{figure}\label{vr2}
\epsscale {0.40}
\plotone{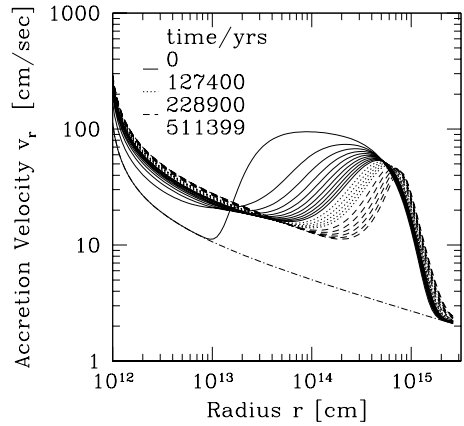}
\caption{Same as Figure 24, but here the radial accretion rate velocity in the disk
is shown. The initial Shakura-Sunyaev solution is given as the dashed-dotted line.}
\end{figure}

\begin{figure}\label{lnu2}
\epsscale {0.40}
\plotone{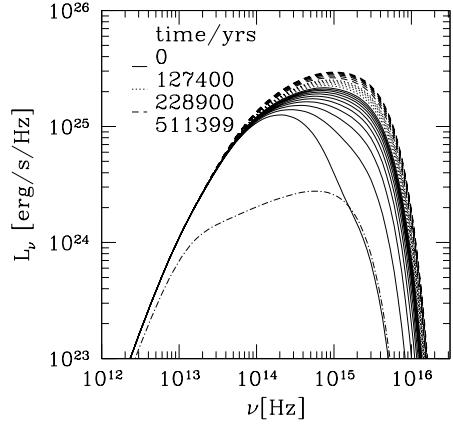}
\caption{The disk spectrum for the case shown in Figure 24.
The initial Shakura-Sunyaev solution corresponds to the dashed-dotted line.}
\end{figure}

\begin{figure}\label{sig1}
\epsscale{0.40}
\plotone{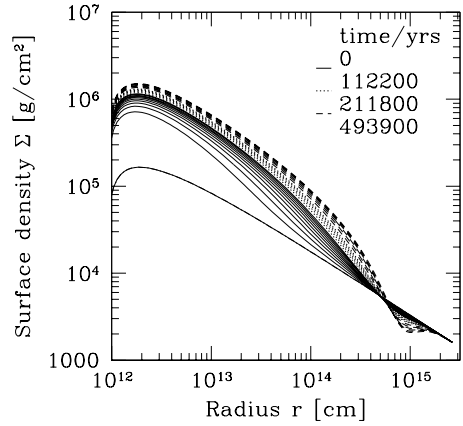}
\caption{Evolution of the surface density of an accretion disk with
wind infall when the direction of the ambient flow is parallel to 
the disk's rotation axis. The disk is a standard
$\alpha$-disk with $\alpha=10^{-4}$ and $\dot M=10^{-6} M_\odot/$yr
around a black hole with mass $10^6 M_\odot$. Different lines correspond to
different time steps (equidistant for equal line styles) as shown in
the legend.}
\end{figure}

\begin{figure}\label{mdot1}
\epsscale{0.40}
\plotone{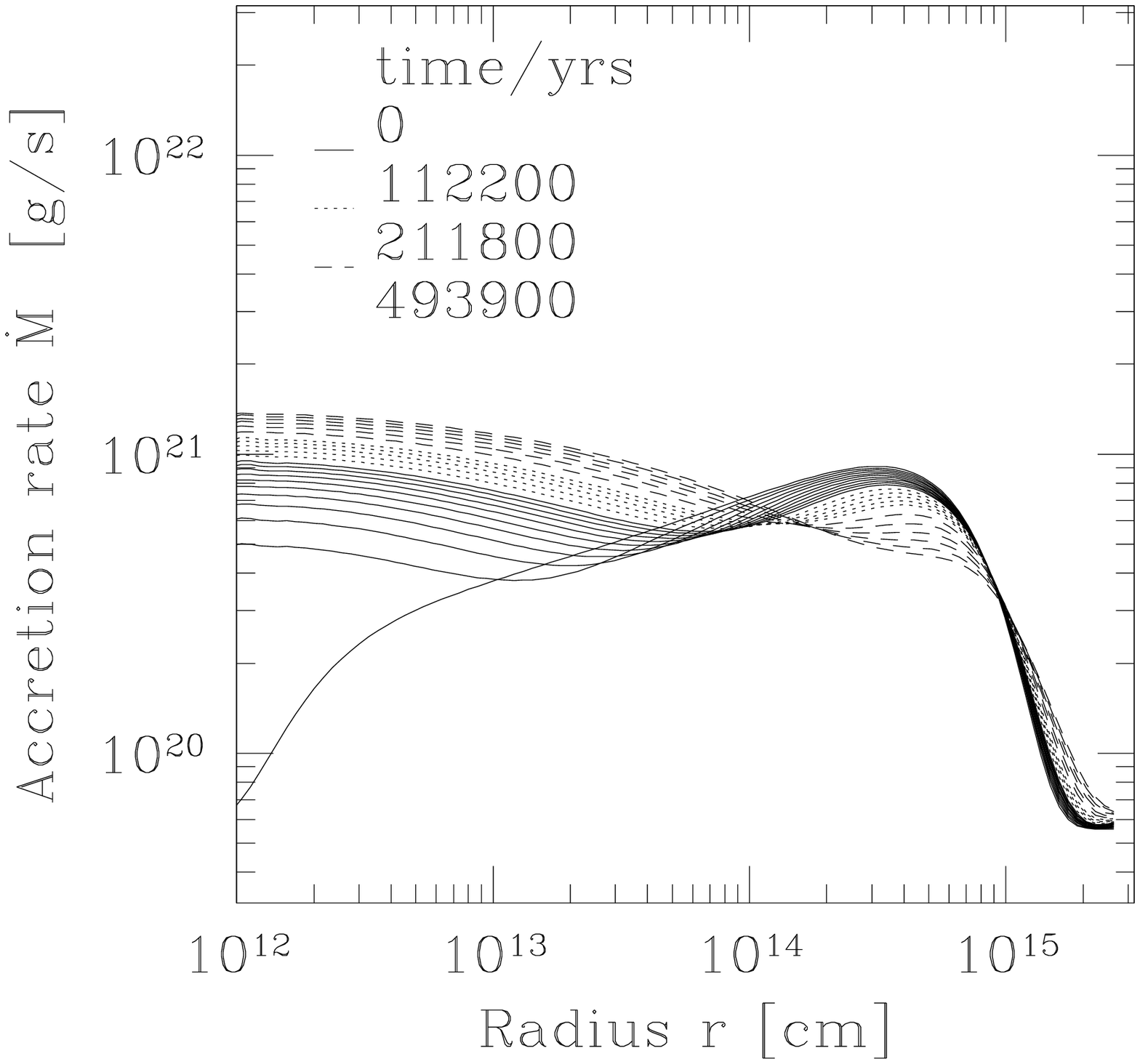}
\caption{Same as Figure 28, but here the local accretion rate in the disk
is shown.}
\end{figure}

\begin{figure}\label{vr1}
\epsscale{0.40}
\plotone{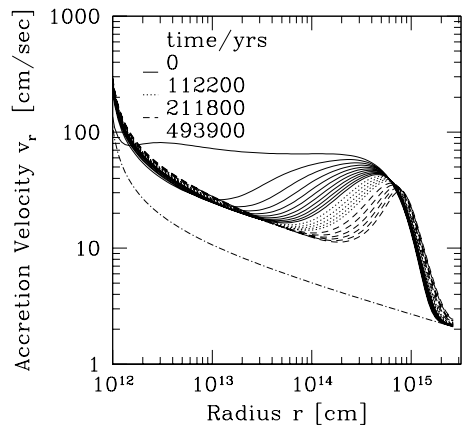}
\caption{Same as Figure 28, but here the radial accretion velocity in the disk
is shown.}
\end{figure}

\begin{figure}\label{lnu1}
\epsscale{0.40}
\plotone{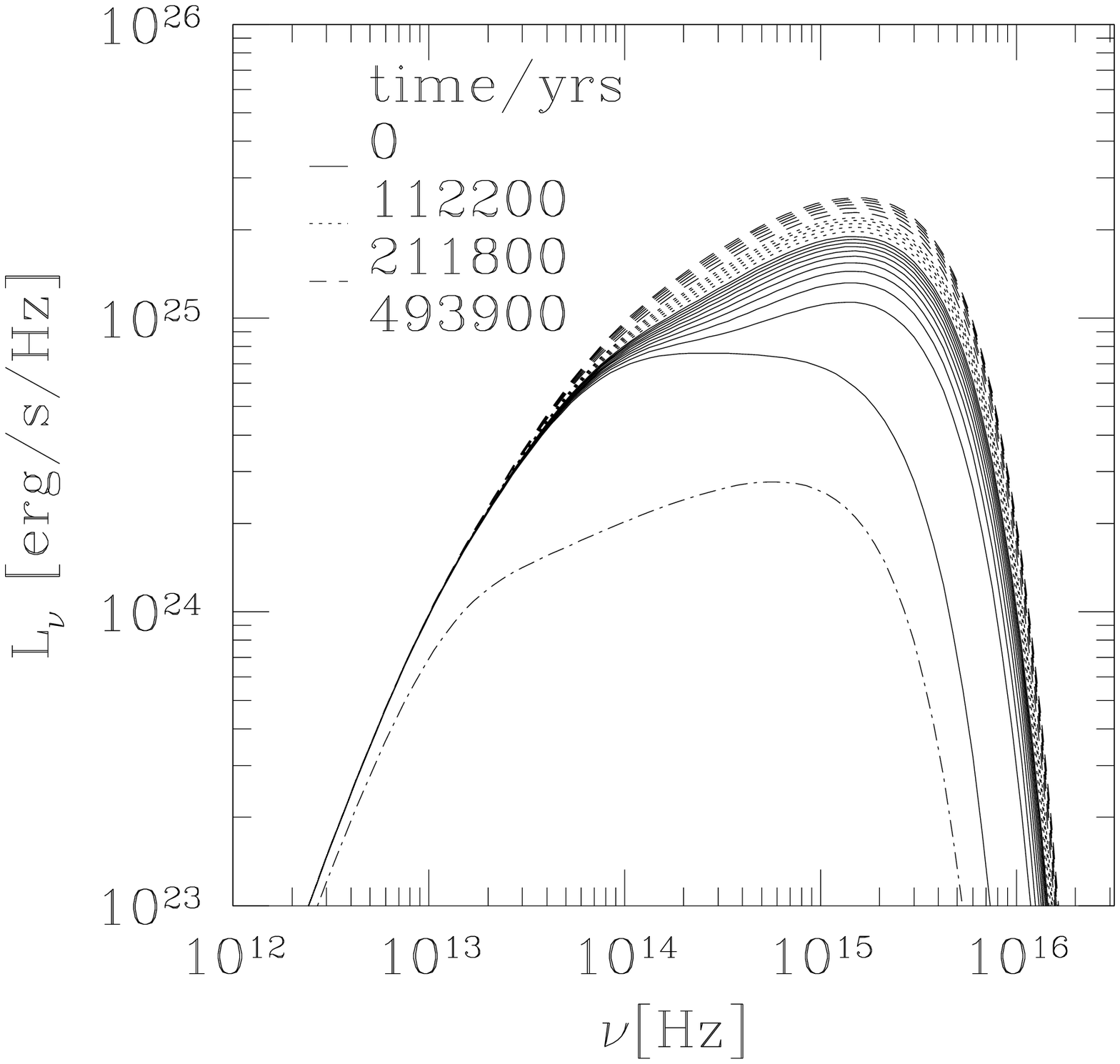}
\caption{The emitted spectrum corresponding to the case shown in
Figure 28.}
\end{figure}

\subsection{The Perpendicular Case}
At large radii, the wind infall dominates the accretion rate. The
infall rate, which is proportional to $\dsigw r^2$, is roughly
constant between the circularization radius and the outermost radius
of the wind. This implies an $\dot M$ that increases linearly
inwards with $r$ in the disk.  In addition, however, the low angular
momentum of the wind will act as a brake to the disk's rotation and
therefore speed-up the radial accretion in the disk. Inside the
circularization radius, where the wind impact is negligible, the
increased mass accretion rate has to be transported by viscous
processes and hence the solution will become more Shakura-Sunyaev like
again. Once the disk has developed towards an equilibrium state the
accretion rate in the disk (Fig.~24) will increase as a power-law
throughout the wind region and become constant inside the
circularization radius.The radial
velocity changes in a commensurate way. 

The effect on the radiated spectrum is predictable: the impact of the
low-angular momentum wind on the fossil disk leads to an immediate
release of energy at large radii where most of the mass is
deposited. This is reflected by the presence of a strong NIR/IR-bump
in the spectrum. Only after a period of a few times $10^5$ years do
the viscous processes begin to transport the increased mass deposition
inwards, leading to a corresponding rise in the spectrum at higher
frequencies. The exact location of the NIR/IR-bump will depend on the
size of the fossil disk. A larger disk will produce a bump at somewhat
lower frequencies and require accordingly a longer time scale to
evolve.

\subsection{The Parallel Case}
A slightly different behavior is evident when the ambient flow is
parallel to the disk axis.  The radial dependence of \dsigw\ has a
steep component (from the top side) like in the perpendicular case,
which, however, has a very small circularization radius. This means
that a significant amount of mass is deposited relatively close
in. Moreover, on the bottom side of the disk, most of the mass is
still being deposited at large radii. Consequently, the whole disk,
and not only the outer part, is forced to react to the infalling wind
immediately, and initially $\dot M$ will be high at all radii. This
increase of $\dot M$ at small radii will lead to an increased
radiation at higher frequencies, while the impacting wind at larger
radii will produce a strong IR/NIR-bump. Since the dissipation of
energy in the wind impact zone is basically happening on a very short
time scale (disk height divided by free-fall velocity), the effect on
the low-frequency part of the spectrum is still more pronounced. Only
after a viscous timescale, when equilibrium is re-established and the
increased mass inflow at large radii is transported through the disk
to the smallest radii, will the spectrum become dominated by the hot,
innermost part of the disk.

\subsection{Precession and Disruption}
A serious limitation of our calculations is that the disk orientation
is fixed in space. However, as we have shown earlier, the impact of
the wind in the parallel case is different on the top and bottom
sides of the disk, and at different azimuths. One may therefore
reasonably expect that this gradient in wind pressure can lead
to a (differential) precession of the disk and possibly also to an eventual 
re-orientation of its symmetry axis or to its overall disruption.
While a disruption would require a fairly sophisticated treatment,
taking into account all possible instabilities, a change in the rotation 
axis can be dealt with in a simple, though crude, way.

Since the most symmetric accretion profile of the infalling gas appears
to be associated with the perpendicular case, where the ram pressure of the
wind is most evenly distributed, one can expect that an otherwise
oriented disk would tend to evolve towards this state.
We also can expect that a change of the disk's orientation
will have occurred by the time the accumulated mass has exceeded the
initial mass in the disk. Interestingly, for the parameters chosen here, the
mass in the inner $10^{13}$ cm of the disk is only $10^{-2}M_\odot$, so that
the inner fossil disk in the parallel case will be overwhelmed
after only a few hundred years, while for the outer part of the disk (at
$10^{15}$ cm) the corresponding time is some $10^4$ years. Consequently for
the type of disk used here, the parallel configuration 
will probably never reach a stationary solution and it will
either be disrupted early or it will evolve towards the perpendicular
case. Again, this behaviour will depend very critically on the initial
disk conditions. Decreasing $\alpha$ even further will increase the
disk mass and decrease the relative influence of the
wind. Alternatively, a larger disk would, at least in its outer parts,
be much more stable, e.g., if the fossil disk would extend to
$10^{17}$cm (in which case $M_{\rm disk}\simeq10^3M_\odot$), the time 
scale for disruption would then increase to more than $10^7$ years.

\section{Conclusions and Possible Applications}

The 3D hydrodynamical calculations used here show that the usual
assumption of a hypersonic free-falling wind is not always accurate.  
The presence of multiple wind sources, the gas conditions at infinity, 
the orientation of the fossil disk, and the thermal pressure of the 
gas all serve to modify the velocity profile of the accreting wind.

The wind infall onto the disk can in principle deposit mass at a
rate anywhere between $\dsigw=$ constant and $\dsigw\propto
r^{-3}$.  The former results from a relatively uniform, cylindrical
flow onto the disk, whereas the latter limit occurs when the inflow
is purely radial.  At least for the two cases considered here,
the actual calculated rate ($\sim r^{-1.8}$) was somewhat closer to the second 
possibility than the first, implying that even though the plasma
velocity configuration at large radii may be uniform, the inflow
becomes more radial at small radii where it merges onto
the disk.  The net effect of this is that $\dot M$ through the
disk then increases roughly linearly with inverse radius, and
the observational impact of a wind infall is then felt on a relatively
short time scale (i.e., hundreds of years) compared to the
viscous time scale of a fossilized disk.  We have also considered
the likely effect on the disk's stability due to this infall, and
based on our results, we concluded that a disk will probably
rotate in time until its symmetry axis is perpendicular to the
incoming wind flow direction. 

An additional rather robust result of our simulations is that the
emitted spectrum of a wind-fed disk can be observationally distinct
from that of a pure Shakura-Sunyaev flow.  Depending on where most
of the wind deposition occurs, the ensuing dissipation of the wind's
kinetic energy at the disk's surface contributes a significant
fraction of the overall infrared luminosity.  For example,
in Figure 31, not only is the flux higher by an order of magnitude
compared to that of the Shakura-Sunyaev solution, but more importantly
as an observational diagnostic, the spectrum at infrared
frequencies is steeper than in the Shakura-Sunyaev case.  This
feature is relatively independent of any particular application
and so it is likely to be common in any wind-fed disk system.

We have considered the possibility that Sgr A* may be surrounded by a
fossil disk accreting at a rate of $10^{-6} M_\odot/$yr$^{-1}$, while 
intercepting an infalling wind of $10^{-4} M_\odot/$yr$^{-1}$.
Current infrared observational limits for Sgr A* are very restrictive, with
$L_\nu \la 10^{21}$ ergs sec$^{-1}$Hz$^{-1}$ at $10^{14}$ Hz.  We have found
that, for the wind configuration we have considered here, the predicted 
infrared luminosity is more than 3 orders of magnitude greater than the 
observational limits.  It seems likely that even if the gas circularizes
at a large radius ($\ga10^{16}$ cm or $\lambda \ga 100$)
and has an extremely low viscosity ($\alpha\la10^{-4}$),
the observational limits will be violated.  In any case, 
hydrodynamical simulations
(\cite{CM97}) suggest that, even including more realistic individual
point sources, the value of $\lambda$ is probably $\la 50$.
Thus it seems unlikely that a standard massive fossilized
disk may be present around Sgr A*.

Many of these conclusions apply also to other types of
disks, such as an advection dominated one.  However, the
equations we have used in this paper to describe the
disk evolution cannot adequately handle the internal structure
of such a configuration, and we must await future developments
employing the proper global equations in order to make
conclusive statements involving the fate of these disks when
they intercept a rather strong Bondi-Hoyle inflow.
The state variables depend on the gradients of quantities 
such as the temperature, as well as their local values. Nonetheless,
it is safe to conclude that the kinetic energy dissipated
above the disk by the infalling wind is likely to produce
a spectral signature that is inconsistent with the observations
for any type of large scale disk.  

Thus, based on our application to the Galactic center
source, Sgr A*, our results would seem to argue against there 
being any type of large scale disk (i.e., with radius $\gg 50 r_g$)
feeding the central mass concentration, and that instead
the inflow into this object probably swirls toward small
radii without ever forming a stable, flattened configuration. 

{\bf Acknowledgments} This research was partially supported by NASA
under grant NAGW-2518.

{}
\end{document}